# SPATIALLY RESOLVED LOCK-IN MICRO-THERMOGRAPHY (SR-LIT): A TENSOR ANALYSIS-ENHANCED METHOD FOR ANISOTROPIC THERMAL CHARACTERIZATION

**Dihui Wang[1], Heng Ban[1,*], Puqing Jiang[2,*]**

[1]Swanson School of Engineering of University of Pittsburgh, Pittsburgh, US

[2]School of Power and Energy Engineering, Huazhong University of Science and Technology, Wuhan, Hubei 430074, China

*Corresponding authors: hengban@pitt.edu (H. B.); jpq2021@hust.edu.cn (P. J.)

## *Abstract*

While high-throughput (HT) computations have streamlined the discovery of promising new materials, experimental characterization remains challenging and time-consuming. One significant bottleneck is the lack of an HT thermal characterization technique capable of analyzing advanced materials exhibiting varying surface roughness and in-plane anisotropy. To tackle these challenges, we introduce spatially resolved lock-in micro-thermography (SR-LIT), an innovative technique enhanced by tensor analysis for optical thermal characterization. Our comprehensive analysis and experimental findings showcase notable advancements: We present a novel tensor-based methodology that surpasses the limitations of vector-based analysis prevalent in existing techniques, significantly enhancing the characterization of arbitrary in-plane anisotropic thermal conductivity tensors. On the instrumental side, we introduce a straightforward camera-based detection system that, when combined with the tensor-based methodology, enables HT thermal measurements. This technique requires minimal sample preparation and enables the determination of the entire in-plane thermal conductivity tensor with a single data acquisition lasting under 40 seconds, demonstrating a time efficiency over 90 times superior to state-of-the-art HT thermology. Additionally, our method accommodates millimeter-sized samples with poor surface finish, tolerating surface roughness up to 3.5 μm. These features highlight an innovative approach to realizing HT and accurate thermal characterization across various research areas and real-world applications.

## 1. INTRODUCTION

The rapid progress in uncovering advanced materials has captured significant attention for its potential to overcome constraints in emerging technologies[1–3]. To achieve this goal, researchers have extensively pursued three primary approaches: combinatorial synthesis, computation, and property characterization. The rapid development of initiatives such as the Material Project and Google DeepMind has facilitated high-throughput (HT) methods to predict, screen, and optimize materials at an unprecedented scale and pace[4–7]. However, the effectiveness of these data-driven algorithms heavily relies on large and accurate datasets of material properties[8–10]. Consequently, HT characterization becomes a crucial step in this integrated cycle.

However, the characterization of thermal transport properties, notably thermal conductivity ($k$), faces challenges due to the lack of HT characterization techniques[11,12]. To date, only a few hundred out of approximately 100,000 laboratory-synthesized materials listed in the Inorganic Crystal Structure Database have experimentally measured $k$ values[13–15], highlighting a significant bottleneck. This challenge is further exacerbated by the increasing number of anisotropic materials, especially those exhibiting in-plane anisotropy, which offer direction-dependent properties[16] crucial for various applications such as electronics[17–21], thermal management[22,23], and photonics[24,25]. Limited knowledge of thermal transport properties not only restricts the optimized use of existing materials but also impedes data-driven material discovery efforts[11,12].

An ideal HT thermal characterization method should feature simple sample preparation, rapid data acquisition, and accurate measurements, even for challenging arbitrary in-plane anisotropic thermal conductivity tensors. However, current techniques fall short of meeting all three criteria. Contact methods like modified $3\omega$ methods, though employed for anisotropic thermal conductivity tensor measurement [26–31], suffer from delicate sample preparation and lack spatial resolution, limiting their efficiency and applicability. Non-contact methods, on the other hand, offer simpler sample preparation and spatially resolvable thermal property measurements[32–34], making them attractive for integration into autonomous platforms such as addictive manufacturing for in-situ monitoring[35] and autonomous labs for HT characterization[10,36].

The pump-probe method, a common non-contact measurement technique, has witnessed various advancements such as elliptical-beam time-domain thermoreflectance (TDTR)[37,38], beam-offset frequency-domain thermoreflectance (BO-FDTR)[39], and spatial-domain thermoreflectance (SDTR)[40–42], for measuring anisotropic in-plane thermal conductivities. However, these pump-probe techniques face limitations when measuring in-plane anisotropic thermal conductivity tensors. They rely on a sequential probing system, which is inherently time-consuming due to repetitive data acquisition. Moreover, they operate at high modulation frequencies (>10 kHz) [43–46], making measurements highly sensitive to cross-plane rather than in-plane thermal conductivity[47,48]. Additionally, spectral reflection requirements for the probe beam typically demand mirror-smooth sample surfaces[49], limiting real-world applications.

Camera-detection-based thermal characterization techniques, like transient infrared thermography with an IR camera[50–53] or thermoreflectance imaging with a CMOS camera[13], offer HT data acquisition through thermal imaging. Unlike sequential probing, these systems record two-dimensional (2D) surface temperature responses, allowing simultaneous temperature acquisition across a large area of the sample surface[55–61]. Nevertheless, the fundamental limitation of all existing techniques for characterizing in-plane thermal conductivity tensors lies not in the instrumentation but in the methodology, namely the reliance on vector-based analysis.

Vector-based analysis assumes that the signal in one direction (represented as a vector) depends solely on the thermal conductivity vector along that specific direction. However, for materials with in-plane anisotropy, a temperature gradient in one direction can result in heat conduction in another direction. Ignoring these intricate interplays can lead to an incomplete or misleading interpretation of the anisotropic thermal conductivity tensor. To enhance the precision of vector-based methods, an iterative process that includes data in multiple directions has been developed[37,38,40]. However, this process compromises efficiency and still struggles to accurately measure materials with unknown crystalline orientation[39,40,62]. An ideal methodology would employ a tensor-based analysis, utilizing 2D data maps instead of selected directional 1D data lines to accurately capture the inherent 2D nature of in-plane anisotropic thermal conductivity tensors. Integrating tensor-based methodology with a camera-based experimental system could pave the way for next-generation thermology.

Here, we propose spatially-resolved lock-in IR micro-thermography (SR-LIT), a tensor analysis-enhanced thermal characterization method. This method features straightforward, all-optical instrumentation, utilizing an electronically modulated, fiber-coupled laser as the heat source and an IR micro-thermography system to map surface temperature at micron-scale resolution. Integrated with the novel tensor-based methodology, SR-LIT achieves HT measurement of the in-plane anisotropic thermal conductivity tensor, with a total data acquisition time of under 40 seconds. For comparison, state-of-the-art thermology methods require an hour to conduct measurements in four directions, making SR-LIT over 90 times more efficient. Through detailed analysis and demonstrative measurements, we show that the tensor-based analysis addresses the fundamental limitation of vector-based analysis by fully exploiting the spatially resolved thermal response. This method yields substantially improved measurement quality over current vector-based methods for the in-plane anisotropic thermal conductivity tensor. Furthermore, we demonstrate that SR-LIT tolerates samples with poor surface finish (up to $r_a = 3$ μm), exhibiting superior versatility compared to thermoreflectance and 3ω techniques. Given these attributes, both the experiment setup of SR-LIT and the tensor methodology offer new avenues for HT thermal characterization across manufacturing facilities and research laboratories.

## 2. INSTRUMENTATION

**Figure 1** (a) depicts the schematic diagram of SR-LIT. A continuous wave (CW) TEM00 fiber-coupled laser at 785 nm wavelength (Thorlabs S4FC785) is directly modulated by a function generator to provide an intensity-modulated heating source at a frequency $f_0$. The output of the fiber-coupled laser is first collimated and then focused through a plano-convex lens onto the sample surface at an incident angle of approximately $30°$. This non-zero incident angle arises due to the short working distance of the IR microscope objective lens, resulting in a slightly elliptical laser spot on the sample surface. Nonetheless, this elliptical laser spot can be accurately characterized and does not impact the measurements, as detailed in Supplementary Material. Sec. S1.

On the detection side, the mid-wavelength IR (MWIR) radiation ranging from $2.4$ to $5$ μm emitted from the surface is collected by an IR micro-thermography system (QFI InfraScope MWIR Temperature Mapping Microscope). This system is equipped with a liquid nitrogen-cooled InSb focal plane array (FPA) and operates at a framerate of $f_{\text{sampling}} = 53.4$ Hz, allowing the capture of up to 2000 frames in each measurement. The field of view (FOV) of the microscope system depends on both the pixel size of the FPA and the chosen objective lens. With a pixel size of 24 μm and a resolution of $500 \times 500$, and employing a typically used $4\times$ objective lens, the pixel resolution (pixel size in FPA/magnification) is 6 μm, yielding a corresponding FOV of $3000 \times 3000$ μm$^2$. To ensure consistent absorption of the laser beam and detection of the IR emission, all samples are coated with a standardized 100 nm Ti transducer using an E-beam evaporation system (Plassys Electron Beam Evaporator MEB550S). Exemplary temperature maps are provided in **Figure 1** (b), wherein the temperature is depicted as a function of spatial coordinates and time $T_s(x,y,t)$ (**Figure 1** (c)).

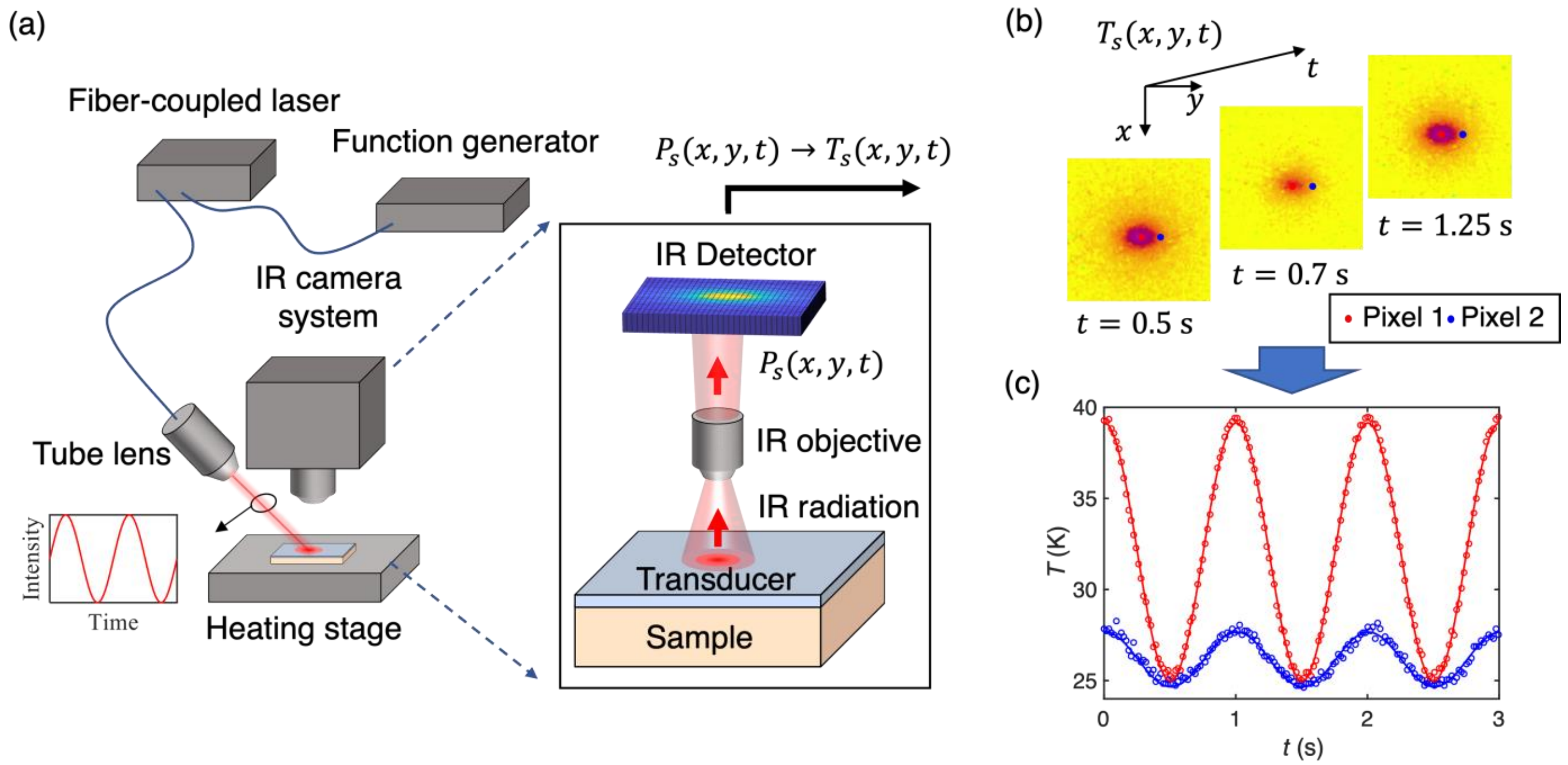

**Figure 1.** (a) Schematic of SR-LIT: A fiber-coupled laser is modulated to periodically heat the sample, while the surface-emitted IR radiation $P_s(x,y,t)$ is collected by an IR micro-thermography system. (b) Exemplary surface temperature map $T_s(x,y,t)$ showing an elliptical heating pattern at 3 different time steps. (c) For each pixel, the temperature is recorded as a function of time, where the open circles represent the measured data, and the solid lines depict the sinusoidal wave fitted by FFT.

## 3. METHODOLOGY

### 3.1 Experiment procedure

The experimental procedure and the tensor-based methodology are illustrated through the measurement of a standard isotropic sample, fused silica. For sample preparation, the sample surface undergoes cleaning via ion etching and is then coated with a 100-nm-thick Ti transducer. The thickness of the Ti transducer is verified using a surface profilometer (Bruker DektaXT). The thermal conductivity of the Ti layer is determined to be approximately $15\ \mathrm{Wm^{-1}K^{-1}}$ by applying the Wiedemann-Franz law, derived from its electrical resistivity measured by a four-point probe.

The sample is subjected to heating by a sinusoidally modulated pump beam with a frequency of $f_0 =$ 5 Hz. Several critical considerations affect the choice of $f_0$. Firstly, the in-plane thermal diffusion length induced by the heating, denoted as $d_r = \sqrt{k_r/\pi f_0 C}$, must be at least three times the $1/e^2$ radius of the laser spot, thus enhancing sensitivity to in-plane thermal transport[40]. Secondly, $f_0$ should be less than the Nyquist frequency of 27 Hz (half of the camera's sampling rate) to prevent any distortion or leakage in the signal. Lastly, the frequency must be high enough to circumvent the predominance of pink noise ($1/f$ noise).

The IR micro-thermography system monitors the surface temperature map $T_s(x, y, t)$ for thermal analysis. Temperature determination is facilitated by a pre-calibrated surface emissivity map $\varepsilon_s(x, y)$ (see the measured emissivity map in Supplementary Material. Sec. S2). To streamline mathematical representations, the long and short radii of the elliptical heating spot are aligned with the x- and y-axes of the coordinate system. This alignment obviates the need to account for a tilted ellipse as the heating source (see the calibration of alignment in Supplementary Material. Sec. S3).

### 3.2 Tensor-based analysis

The tensor-based analysis commences with an examination of the spatially resolved temperature response. The initial step involves transforming the measured time-domain surface temperature map $T_s(x, y, t)$ into a frequency-domain map $\Theta_s(x, y, \omega_0)$ using Fast Fourier Transform (FFT). To minimize spectral leakage and scalloping loss, careful consideration is given to the choice of total sampling time [63,64], which we found to be optimal at $t_{\mathrm{samp}} = 37\ \mathrm{s}$ for our camera system (see detailed FFT procedure in Supplementary Material. Sec. S4).

The FFT process yields four frequency-domain maps: the AC amplitude map, DC amplitude map, phase map, and SNR map (see **Figure 2**). Each map serves a distinct purpose in the fitting process. The AC amplitude map (**Figure 2** (a)) is used for the spot size fitting, thereby simplifying measurements by eliminating the need for separate spot size measurements and enhancing reliability by mitigating unintentional spot size variations between different measurement sets. The capability to measure the steady-state (DC) amplitude map (**Figure 2** (b)) offers a more comprehensive understanding of the modulated heating event than most lock-in methods that neglect the DC response. Overlooking the DC amplitude can lead to a significant underestimation of the overall temperature rise[65]. Subsequently, the phase map (**Figure 2** (c)) constitutes critical data, which is then compared with a 3D multi-layer heat diffusion model for thermal analysis (refer to the mathematical model in Appendix. S1). Lastly, the signal-to-noise ratio (SNR) map allows for a visual representation of the spatially resolved SNR (**Figure 2** (d)), with pixels exhibiting a low SNR (<10) filtered out to enhance the fitting quality.

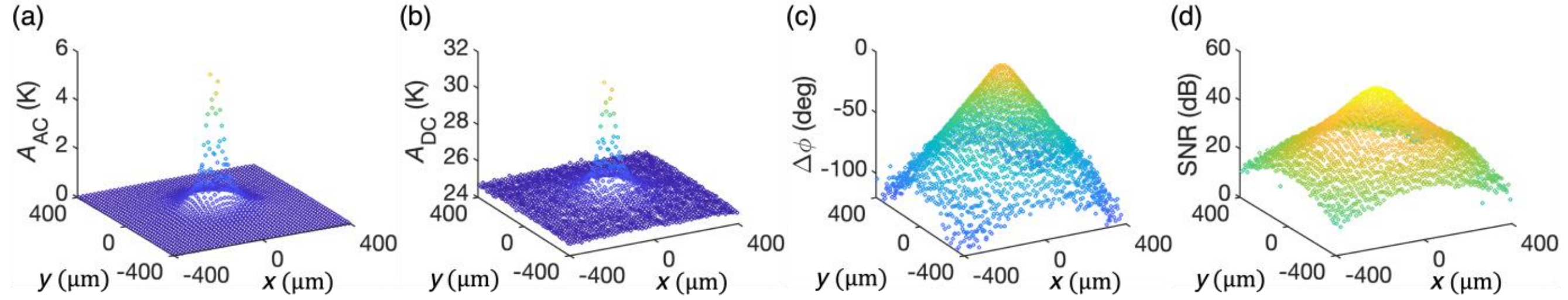

**Figure 2.** (a) Exemplary data maps after FFT processing: (a) modulated (AC) amplitude map, (b) steady-state (DC) amplitude map, (c) phase map, and (d) Signal-to-Noise Ratio (SNR) map. This dataset is obtained from fused silica coated with a 100-nm-thick Ti transducer.

### 3.2.1 1D sensitivity analysis

Sensitivity analysis along the x- and y-axes serves as the initial guide for utilizing $A_{\mathrm{norm}}$ and $\Delta\phi$ maps for property fitting (see the definition of sensitivity coefficient in Appendix. S2). $A_{\mathrm{norm}}$ represents the amplitude normalized by the maximum amplitude at the heating center, while $\Delta\phi$ is the differential phase, $\Delta\phi = \phi - \phi_c$, with $\phi_c$ being the phase measured at the heating center. Normalizing the amplitude eliminates the need for calibrating the laser power and the optical absorptivity of the metallic transducer. Meanwhile, differencing the phase removes any phase shift introduced by the electronic components of the system[40]. Moreover, the original phase $\phi$ tends to exhibit high sensitivity to cross-plane heat transport[32,53], which can be significantly suppressed by focusing on the differential phase $\Delta\phi$.

The determination of laser spot size is facilitated by analyzing $A_{\mathrm{norm}}$. **Figure 3** (a) and (b) reveal that in the near-center region ($x_c < 2w$, where $w$ is the $1/\mathrm{e}^2$ radius), $A_{\mathrm{norm}}$ along the x- and y-directions is dominantly sensitive to $w_x$ and $w_y$, respectively. This observation holds true across various tested property combinations (nominal values of input parameters can be found in Supplementary Material. Sec. S5). As mentioned earlier, this feature enables the fitting of spot radii for each measurement, thus enhancing measurement reliability by mitigating unintentional spot size variations between different measurement sets.

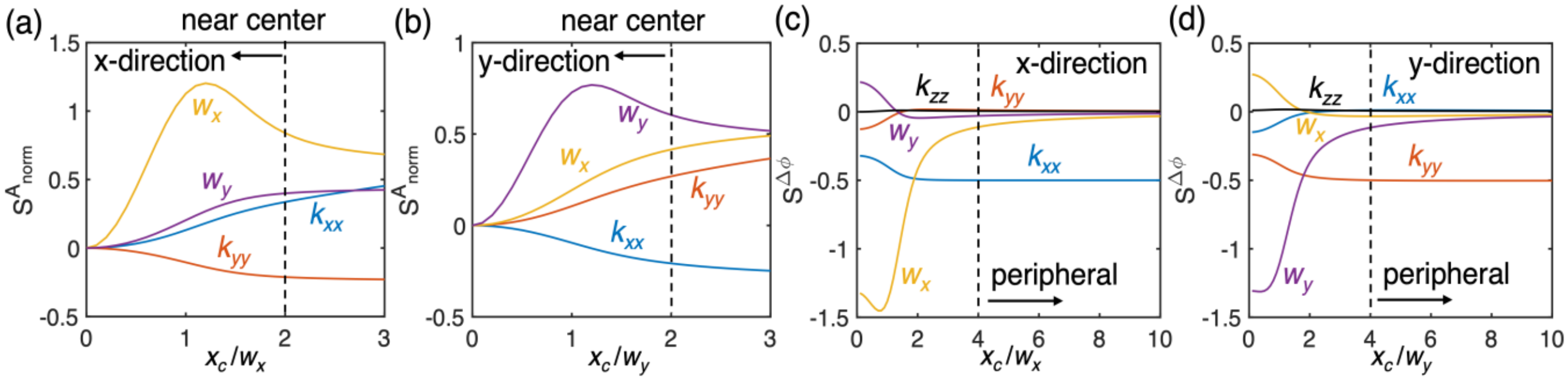

**Figure 3.** Sensitivities of $A_{\mathrm{norm}}$ and $\Delta\phi$ as a function of normalized offset distance (the offset distance $x_c$ divided by the spot size $w$ along the offset direction) in (a, c) x-direction and (b, d) y-direction, respectively. This sensitivity analysis is based on fused silica. Parameters with near-zero sensitivities, such as thermal boundary conductance and metallic transducer properties, are omitted from the plots for the sake of clarity.

The determination of the in-plane thermal conductivity tensor is facilitated by analyzing $\Delta\phi$. **Figure 3** (c) and (d) show that in the peripheral region ($x_c > 4w$), $\Delta\phi$ is predominantly sensitive to $k_{\parallel}$ (the in-plane thermal conductivity along the offset direction, which is $k_{xx}$ for x-direction scan and $k_{yy}$ for y-direction scan). Therefore, the in-plane thermal conductivity of the sample can be accurately determined by fitting the $\Delta\phi$ signals at large $x_c$. However, the SNR decreases as the offset distance increases. To better utilize the $\Delta\phi$ signals at large $x_c$, it is essential to minimize random noise. This can be achieved by employing a $5 \times 5$ average filter (details regarding the average filter and noise reduction can be found in Supplementary Material. Sec. S6).

### 3.2.2 2D sensitivity map

To visualize how the 2D data map enhances the measurement of the in-plane thermal conductivity tensor, we progress from analyzing the 1D sensitivity line to a 2D sensitivity map. Interestingly, our observations indicate that $\Delta\phi$ is highly sensitive to $k_{xx}$ not only along the x-axis but also within an angular range $\alpha_x$, as depicted in **Figure 4** (a). Similarly, $\Delta\phi$ exhibits high sensitivity to $k_{yy}$ along the y-axis direction within an angle of $\alpha_y$ (**Figure 4** (b)). These findings suggest that utilizing data beyond just the *x*- and *y*-axes, as is common in vector-based methods, is advantageous for determining the in-plane thermal conductivity tensor, even for isotropic materials.

For optimal fitting quality, the $\Delta\phi$ map requires filtering. Data points in the near-center region, where $x_c < 4w(\theta)$, are excluded due to their low sensitivity to the in-plane thermal conductivity. The radius $w(\theta) = \frac{w_x w_y}{\sqrt{w_x^2 \sin^2\theta + w_y^2 \cos^2\theta}}$ is angle-dependent, reflecting the elliptical shape of the heating pattern. Additionally, data points with an SNR of less than 10 are removed. After the filtering process, the resulting data map appears as a ring shape, consisting of 629 data points (**Figure 4** (c)). In comparison, data lines along the x- and y-axes consist of a total of 21 data points. Therefore, the resulting map spans approximately 30 directions. This broad coverage provides a substantially more comprehensive depiction of in-plane heat transport than conventional vector-based methods. Finally, the filtered 2D data map is used to fit the in-plane thermal conductivity tensor, $\boldsymbol{k} = \begin{bmatrix} k_{xx} & k_{xy} \\ k_{yx} & k_{yy} \end{bmatrix}$. Since $k_{xy} = k_{yx}$ based on the Onsager relation[66,67], only $k_{xx}$, $k_{yy}$, and $k_{xy}$ ($k_{xy} = 0$ for isotropic materials) are treated as the unknown parameters and are fitted using a nonlinear regression method[68–70].

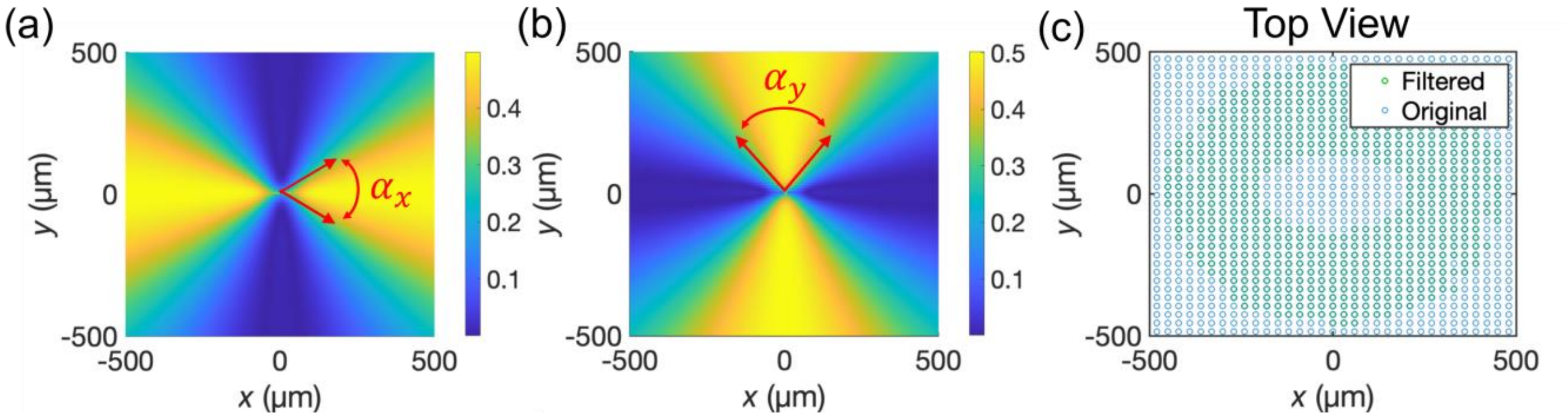

**Figure 4.** Sensitivity map for (a) $k_{xx}$ and (b) $k_{yy}$ of fused silica with high sensitivity region $\alpha_x$ and $\alpha_x$ respectively. Demonstrative data map collected on a fused silica sample. (c) Top view of the filtered data (green dots) and original measurement data (blue dots) map.

## 4. RESULTS

### 4.1 Isotropic materials

The $\Delta\phi$ map obtained for the fused silica sample is presented in **Figure 5**, alongside the best-fitted $\Delta\phi$ map derived from the model based on the best-fitted parameters: $k_{xx} = 1.48 \pm 0.06\ \mathrm{Wm^{-1}K^{-1}}$ and $k_{yy} = 1.5 \pm 0.06\ \mathrm{Wm^{-1}K^{-1}}$. These results align well with the literature value of $k_{\mathrm{lit}} = 1.3{\sim}1.5\ \mathrm{Wm^{-1}K^{-1}}$[40,71]. For comparison, conventional vector-based analysis was applied to data along the x- and y-directions, yielding $k_{xx} = 1.45 \pm 0.08\ \mathrm{Wm^{-1}K^{-1}}$ and $k_{yy} = 1.48 \pm 0.09\ \mathrm{Wm^{-1}K^{-1}}$, respectively. As expected, the vector-based analysis provides good accuracy for isotropic materials. However, the $\pm 2\sigma$ uncertainty associated with the vector-based analysis is about 30% higher than that of the tensor-based analysis (see Appendix. S3 for the uncertainty propagation formula). This observation underscores the superior performance of the tensor-based analysis, even for in-plane isotropic materials.

We further demonstrate SR-LIT measurements on (0001) sapphire, revealing in-plane thermal conductivities of $k_{xx} = 37.7 \pm 2.7\ \mathrm{Wm^{-1}K^{-1}}$ and $k_{yy} = 37.2 \pm 2.6\ \mathrm{Wm^{-1}K^{-1}}$. These results agree well with the literature range of $34{\sim}38\ \mathrm{Wm^{-1}K^{-1}}$[72–74]. It's noteworthy that the percentage uncertainty of our measurements for sapphire is slightly elevated compared to that of fused silica. This can be attributed to increased noise when measuring high thermal conductivity materials with the limited laser power of our current system. However, this challenge can be easily overcome by employing a laser with higher output power for the measurement and should not be misconstrued as an intrinsic limitation of the technique.

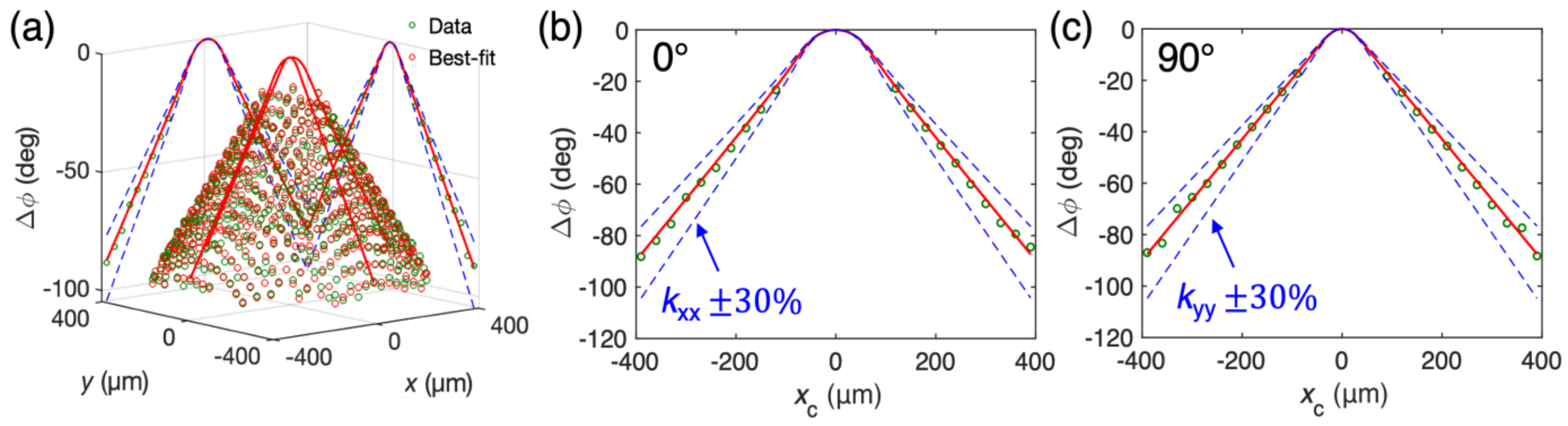

**Figure 5.** Demonstrative SR-LIT fitting map on a fused silica sample. (a) Measured $\Delta\phi$ map (green dots) alongside the best-fitted $\Delta\phi$ map (red dots) and the best-fitted curves (red). The measured data, intersected by the x-z plane and the y-z plane, are projected on the y- and x-plane, respectively. (b) and (c): measured $\Delta\phi$ data (green dots) intersected by planes at selected directions of $0°$ and $90°$, respectively, along with the best-fitted curves (red) and $\pm 30\%$ bounds of the best-fitted thermal conductivity values.

## 4.2 In-plane anisotropic materials

In-plane anisotropic materials exhibit varying thermal properties along different crystallographic directions within the transverse plane[73]. To underscore the imperative of tensor-based analysis for in-plane anisotropic tensor measurement, we commence with a thorough map-based sensitivity analysis. Subsequently, experimental validation follows, encompassing hypothetical samples and x-cut quartz under various conditions. Finally, we explore the adaptability of SR-LIT through measurements conducted on samples with rough surfaces.

### 4.2.1 2D sensitivity map

We analyze a 2D sensitivity map to discern the sensitivity pattern of the in-plane anisotropic thermal conductivity tensor. For instance, we consider a hypothetical tensor $\boldsymbol{k}_{\mathbf{in},\mathbf{1}} = \begin{bmatrix} 13.3 & 3 \\ 3 & 9.8 \end{bmatrix}$ $(\mathrm{Wm^{-1}K^{-1}})$. The positive value of $k_{xy}$ suggests that the temperature gradient in the $x$-direction would induce heat flux in the positive $y$-direction, and vice versa. Consistent with observation in isotropic materials, $\Delta\phi$ exhibits high sensitivity to $k_{xx}$ and $k_{yy}$ mainly along the x- and y-directions, within angles $\alpha_x$ and $\alpha_y$, respectively (**Figure 6** (a) and (b)). However, unlike isotropic thermal conductivity tensor, $\alpha_x$ for anisotropic thermal conductivity tensor is narrower, while $\alpha_y$ is broader. This suggests that the ratio of $k_{xx}$ and $k_{yy}$ influences the region of the data map sensitive to $k_{xx}$ and $k_{yy}$. Moreover, the off-diagonal element $k_{xy}$ shows high sensitivity mainly in a direction pointing to 135°, within an angle of $\alpha_{xy}$ (**Figure 6** (c)). Due to the comparatively high sensitivity levels and the distinct sensitivity patterns for each in-plane thermal conductivity tensor element, simultaneously fitting all three elements becomes feasible.

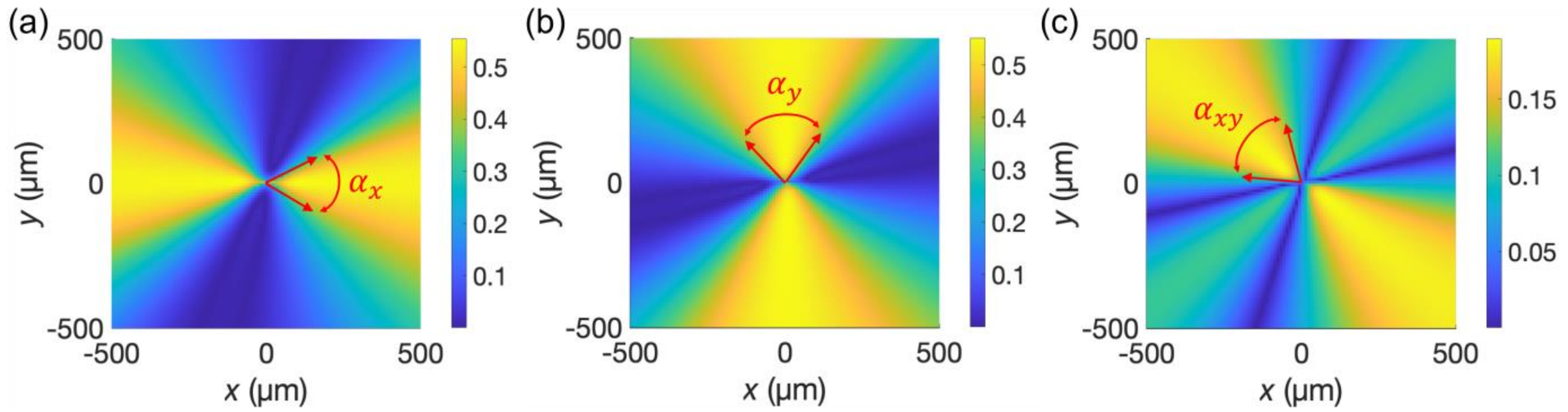

**Figure 6.** Sensitivity maps for $\Delta\phi$ regarding (a) $k_{xx}$, (b) $k_{yy}$, and (c) $k_{xy}$, illustrating the highly sensitive regions represented by $\alpha_x$, $\alpha_y$, and $\alpha_{xy}$, respectively.

It's important to note that the sensitivity map pattern for each element is not fixed. Consider a second case with $\boldsymbol{k}_{\mathbf{in},\mathbf{2}} = \begin{bmatrix} 13.3 & -3 \\ -3 & 9.8 \end{bmatrix}$ $(\mathrm{Wm^{-1}K^{-1}})$. The sensitive region for $k_{xy}$ would now be oriented towards 45°, distinctly different from case 1 (see detailed sensitivity maps in Supplementary Material. Sec. S7). This highlights a critical limitation of current vector-based methods: the selected directions may not always capture the most informative directions for all elements in an arbitrary in-plane thermal conductivity tensor. In contrast, the tensor-based method, encompassing spatially resolved data along all directions, can offer a comprehensive interpretation for arbitrary in-plane anisotropic thermal conductivity tensors.

### 4.2.2 Validation on simulated experiment

We initiate by conducting a simulated experiment to showcase a challenging scenario for the vector-based method, which can, however, be effectively addressed through the tensor-based approach. Here, we provide a hypothetical thermal conductivity tensor $\boldsymbol{k} = \begin{bmatrix} 20 & 2 & 0 \\ 2 & 8 & 0 \\ 0 & 0 & 8 \end{bmatrix}$ $(\mathrm{Wm^{-1}K^{-1}})$. Notably, the off-diagonal element of the tensor is intentionally set to a small but non-zero value. To emulate real experimental conditions during signal generation, we consider the following factors: (1) the sample surface is coated with a 100 nm titanium transducer layer, (2) random noises are introduced at levels comparable to those measured for fused silica, and (3) an elliptical heating pattern with $w_x = 50$ µm and $w_y = 30$ µm.

The simulated 2D data map comprises a total of 1380 data points (refer to **Figure 7** (a) and (b)), equivalent to approximately 49 directions. Analysis of the data map reveals the in-plane thermal conductivity tensor as: $\boldsymbol{k}_{\mathbf{in}} = \begin{bmatrix} 20.05 \pm 1.03 & 2.1 \pm 0.38 \\ 2.1 \pm 0.38 & 8.01 \pm 0.47 \end{bmatrix}$ $(\mathrm{Wm^{-1}K^{-1}})$, closely mirroring the hypothetical tensor, as depicted in the best-fit map in **Figure 7** (c)-(f). For comparative purposes, vector-based analysis is conducted using data along 4 directions at $0°, 45°, 90°,$ and $135°$, resulting in a tensor estimate of $\boldsymbol{k}_{\mathbf{in}} = \begin{bmatrix} 20.46 \pm 2.37 & 1.67 \pm 1.65 \\ 1.67 \pm 1.65 & 7.61 \pm 0.97 \end{bmatrix}$ $(\mathrm{Wm^{-1}K^{-1}})$.

The tensor-based analysis demonstrates superior accuracy and reduced uncertainty compared to the vector-based approach for all tensor elements. Particularly, when using the tensor-based method, the uncertainty for diagonal elements is reduced to half, and for off-diagonal elements, it is reduced to one-fifth of what is observed with the vector-based method. This notable difference arises from the limited signal sensitivity inherent in conventional vector-based methods. As depicted in **Figure 7** (f), even at the optimal direction $(135°)$ for extracting $k_{xy}$, the sensitivity of $\Delta\phi$ to $k_{xy}$ is significantly lower compared to the sensitivity to $k_{xx}$. In contrast, the tensor-based analysis integrates the entire 2D map data and utilizes higher-order statistics to effectively capture the intricate interplay among different tensor elements. Consequently, the tensor-based analysis offers accurate measurements of arbitrary in-plane anisotropic thermal conductivity tensors.

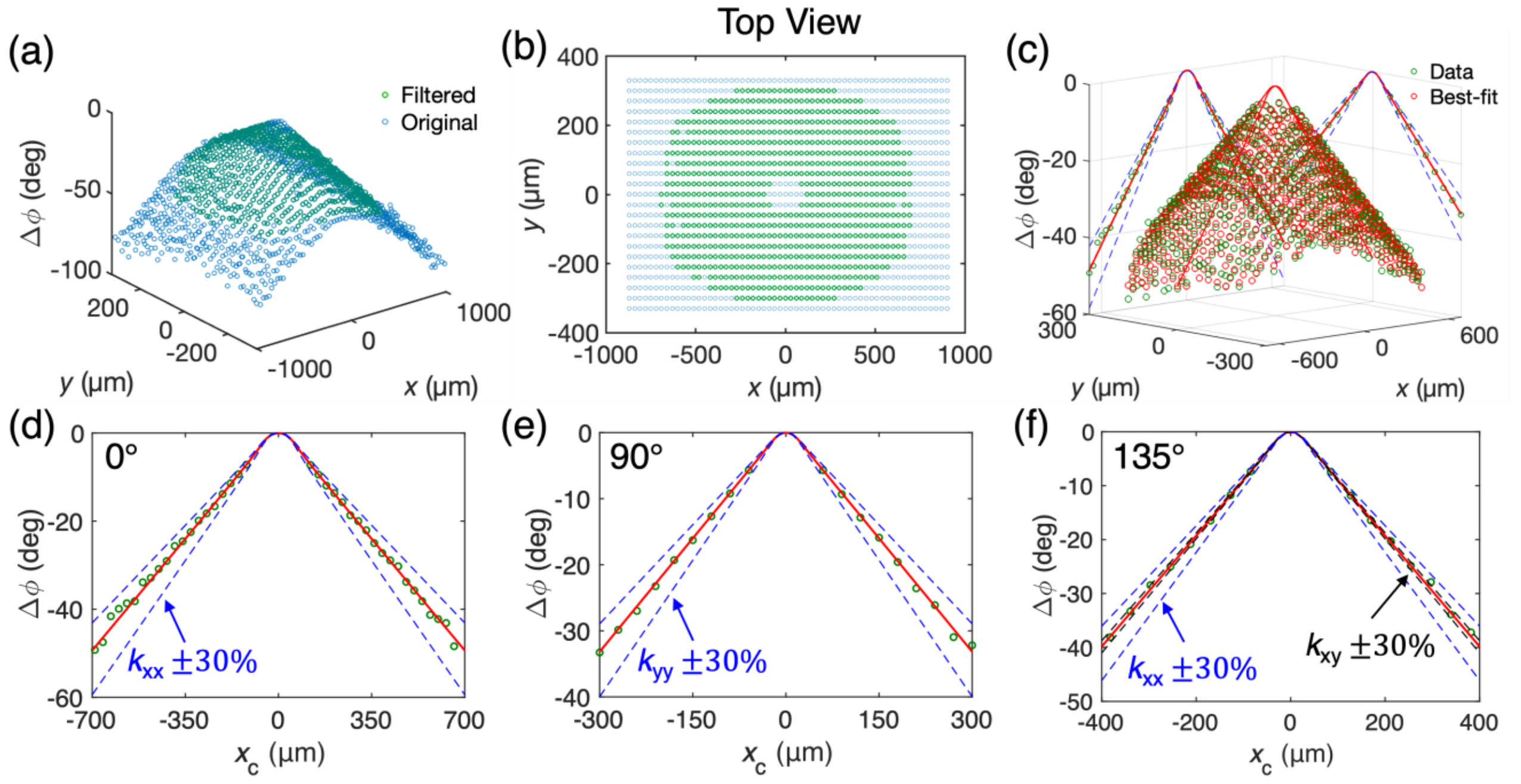

**Figure 7.** Simulated experiment for fitting in-plane anisotropic thermal conductivity. (a, b) Simulated $\Delta\phi$ map (open blue circle) and the filtered ring-shape region (open green circle). Panel (b) provides a top view of (a). (c) Best-fitted map $\Delta\phi$ (open red circle) overlaid with the simulated data. Panels (d)-(f) depict simulated $\Delta\phi$ along three scan directions (0°, 90°, and 135°) with their respective best fitting curves (red) and $\pm 30\%$ bounds of the best-fitted thermal conductivity values.

### 4.2.3 Validation on x-cut quartz

Subsequently, we demonstrate SR-LIT measurements on x-cut ((110)-oriented) quartz. X-cut quartz displays in-plane anisotropic thermal conductivity with $k_c > k_a = k_b$ and features orthogonal in-plane lattice vectors.

In the first part of the demonstration, we orient the sample such that its c-axis aligns parallel to the x-axis of the optical system, effectively setting the target $k_{xy}$ to zero. The resulting filtered ring-shaped data map comprises a total of 1697 data points, representing approximately 31 directions. The best fit of these data yields an in-plane thermal conductivity tensor of $\boldsymbol{k}_{\mathbf{in}} = \begin{bmatrix} 12.1 \pm 0.5 & 0.03 \pm 0.1 \\ 0.03 \pm 0.1 & 7.3 \pm 0.4 \end{bmatrix}$ $(\mathrm{Wm^{-1}K^{-1}})$ (refer to the best-fit map in **Figure 8** (a)). This result exhibits excellent agreement with the literature value $\boldsymbol{k}_{\mathbf{in,lit}} = \begin{bmatrix} 12 & 0 \\ 0 & 6.8 \end{bmatrix}$ $(\mathrm{Wm^{-1}K^{-1}})$[75]. Notably, we refrain from assuming $k_{xy} = 0$ during the fitting process to detect any unintentional rotation of the sample. The fitted result confirms a negligible $k_{xy}$ value.

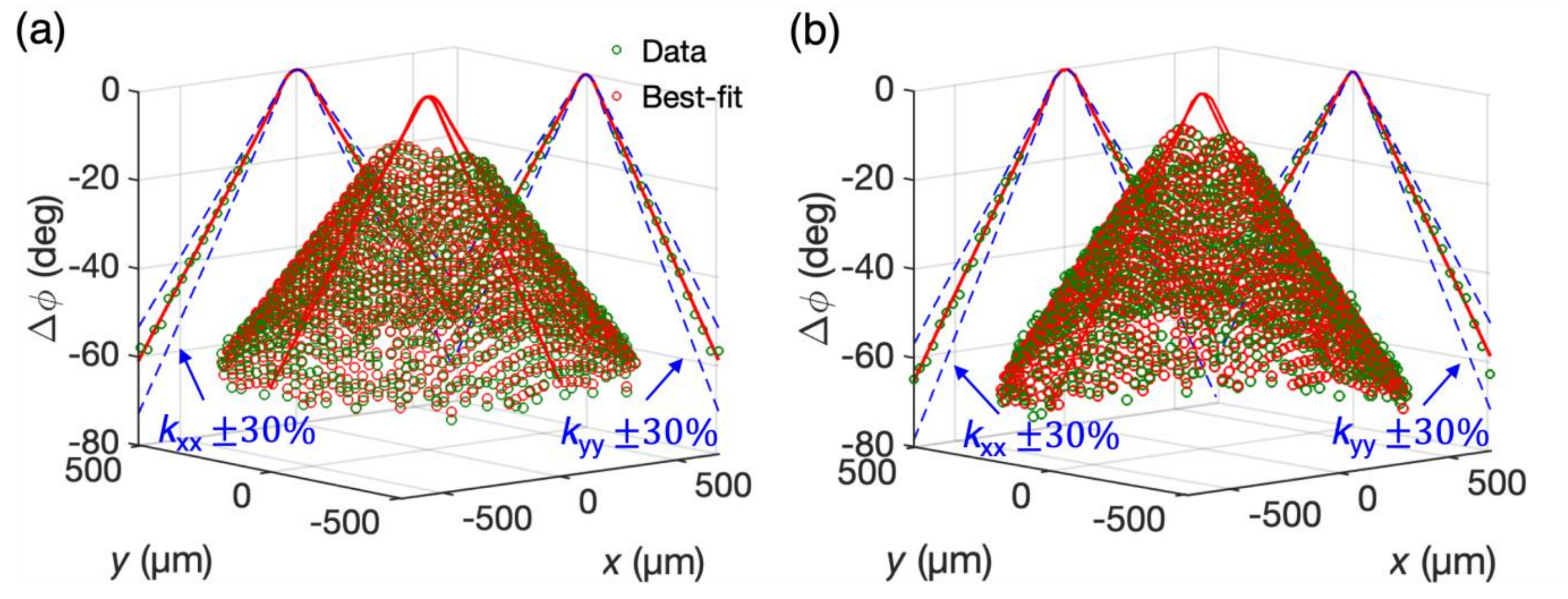

**Figure 8.** Measurements of x-cut quartz under two configurations. (a) The crystalline direction $\vec{c}$ and $\vec{a}$ are aligned parallel to the x- and y-axes of the optical system, respectively. (b) The sample is intentionally rotated $20°$ counterclockwise about the z-axis. The green open circle represents the experimentally measured $\Delta\phi$, while the red open circles are the modeled $\Delta\phi$ based on the best-fitted in-plane thermal conductivity tensor. Data measured along the x- and y-axes are projected on the y- and x-plane, respectively. Blue dashed curves indicating 30% bounds of the best-fitted tensor element along the offset direction ($k_{xx}$ for x-axis data and $k_{yy}$ for y-axis data) are included as guides of sensitivity.

We proceed to evaluate the measurement accuracy following an intentional rotation of the sample by approximately 20°, representing a challenging scenario where determining a small $k_{xy}$ becomes crucial. The measured phase map is plotted in **Figure 8** (b), with the best-fitted $\boldsymbol{k}_{\mathbf{in}} = \begin{bmatrix} 11.2 \pm 0.5 & 1.6 \pm 0.14 \\ 1.6 \pm 0.14 & 7.6 \pm 0.4 \end{bmatrix}$ $(\mathrm{Wm^{-1}K^{-1}})$. With a known $\boldsymbol{k}_{\mathbf{in}}$, the rotation angle $\theta$ can be deduced as $\theta = \frac{1}{2}\arctan\left(\frac{2k_{xy}}{k_{xx}-k_{yy}}\right)$[76,77], resulting in $\theta = 21.5° \pm 1.54°$. This determination agrees well with the designated angle, with the uncertainty falling within a reasonable range. In addition, the crystalline in-plane thermal conductivity tensor is retrieved as $\boldsymbol{k}_{\mathbf{in}} = \begin{bmatrix} 11.8 \pm 0.5 & 0 \\ 0 & 7 \pm 0.5 \end{bmatrix}$ $(\mathrm{Wm^{-1}K^{-1}})$, consistent with the previous measurements and the literature values.

### 4.3 Measurement of rough samples

Conventional pump-probe methods encounter challenges when measuring rough samples, a hurdle efficiently mitigated by utilizing an IR detector. Two critical factors affect the largest tolerable surface roughness: firstly, the depth of field (DoF) of the camera must exceed the surface roughness, and secondly, the thermal diffusion length induced by the modulated heating must far surpass the surface roughness to ensure a uniformly flat sample during thermal measurement[13]. The DoF is contingent upon the specifications of the objective lens and the detection wavelength [78]. Given that IR detection inherently utilizes a longer wavelength than visible light, the corresponding DoF is larger than traditional optical methods using a probe laser in the visible light spectrum. For our current system, utilizing a 4× objective lens and detecting IR radiation with a wavelength of 3.5∼5 μm, the estimated DoF is ~5.8 μm. Coupled with the thermal diffusion length exceeding hundreds of micrometers at a modulation frequency 5 Hz for the tested material, SR-LIT significantly enhances the tolerable surface roughness to a few micrometers. It's worth noting that measuring rougher samples can be achieved by switching to an objective with lower magnification, albeit this might sacrifice some level of spatial resolution.

Both isotropic (soda-lime glass slide) and in-plane anisotropic (x-cut quartz) rough samples are prepared. The samples are roughened using sanding abrasives (Zoro Abrasive) with grit numbers of 24, 36, 60, and 800, where lower grit numbers indicate larger, coarser particles, and vice versa. After grinding, the surface roughness of each sample is determined using a surface profiler (Bruker DektaXT surface profiler), with images of the rough surface profile provided in **Figure 9** (a) and (e).

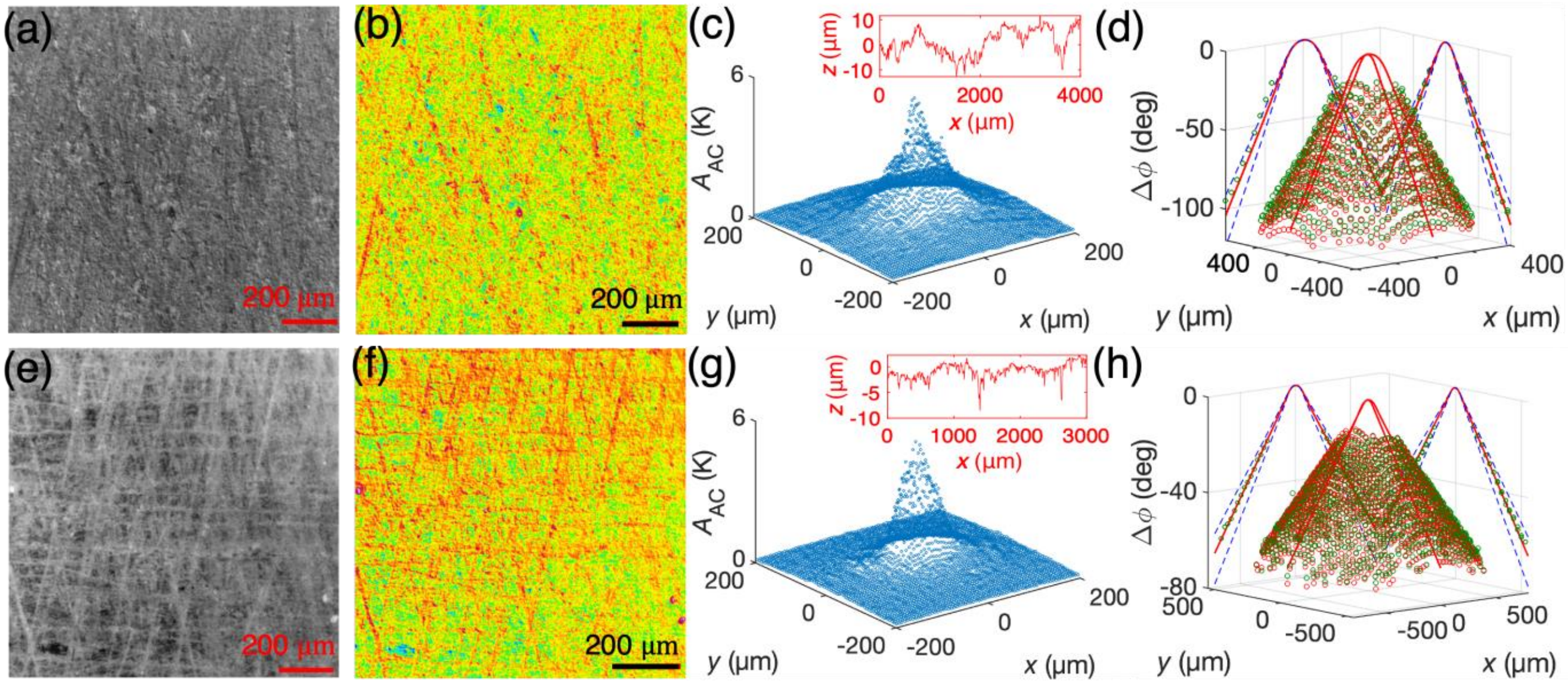

**Figure 9.** Demonstrative plots illustrating SR-LIT measurements of rough samples. (a-d) Measurements of a glass slide with a roughness of $r_a = 3.5$ μm. (e-h) Measurements of an x-cut quartz with a roughness of $r_a = 1.5$ μm. (a, e) Photomicrographs of the glass slide and x-cut quartz sample surfaces. (b, f) Non-uniform emissivity maps . (c, g) AC temperature response in the near-filed region, with inset plots showing surface profiles measured using a stylus profiler. (d, h) Measured and best-fitted phase maps. The red open circles represent the modeled $\Delta\phi$ based on the best-fit in-plane thermal conductivity tensor. Curves indicating 30% bounds of the best-fitted tensor elements along the offset direction ($k_{xx}$ for x-axis data and $k_{yy}$ for y-axis data) are included as guides of sensitivity.

With a roughened surface, the uniformity of surface emissivity is compromised. However, IR micro-thermography can determine the non-uniform emissivity map following the same procedure as detailed in Supplementary Material. Sec. S2[51,79]. Examples of the non-uniform emissivity map of some roughened samples are shown in **Figure 9** (b) and (f).

Large surface roughness could affect the laser profile on the sample surface, causing deviation from a perfect Gaussian distribution (as seen in **Figure 9** (c) and (g) for some examples). However, the non-perfect Gaussian profile does not affect the tensor-based analysis, as only $\Delta\phi$ in the peripheral region, which has negligible sensitivity to the laser spot size, are utilized (e.g., **Figure 3** (c) and (d)). In some extreme cases documented in literatures[80–82], the Gaussian laser beam was even simplified as a point source. The error induced by a non-perfect Gaussian laser profile is estimated to be less than 1% in our SR-LIT measurements (refer to Supplementary Material Sec. S8).

**Figure 10** summarizes the measured in-plane thermal conductivities for rough samples. Soda-lime glass slides, with roughness $r_a$ ranging from 10 nm to 3.5 μm, exhibit consistent results and good agreement with literature values (further details of all measured values are provided in Supplementary Material Sec. S8 and S9). Furthermore, the roughened x-cut quartz sample ($r_a = 2.8$ μm) has been accurately measured, with the best-fit in-plane thermal conductivities of $k_c = 12.3 \pm 0.5\ \mathrm{Wm^{-1}K^{-1}}$ and $k_a = 7.3 \pm 0.3\ \mathrm{Wm^{-1}K^{-1}}$ (refer to **Figure 9** (h)). Importantly, the uncertainty levels for both types of samples are consistent with those observed for smooth samples. These findings highlight the robustness of SR-LIT in accurately measuring samples with rough surfaces.

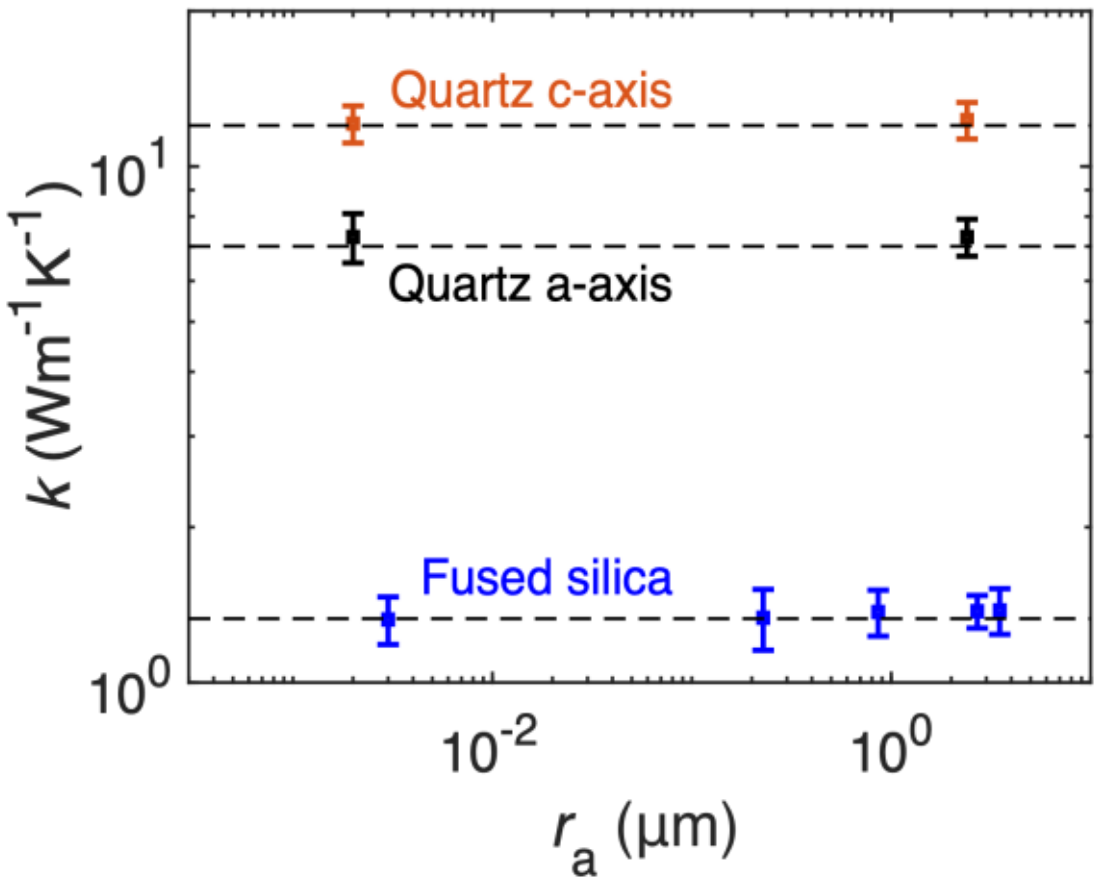

**Figure 10.** SR-LIT measurements of in-plane thermal conductivity for rough samples with varying surface roughness. The dashed black lines are provided as visual guides.

## 5. SUMMARY AND OUTLOOK

All measured materials are summarized in **Figure 11**. This study introduces SR-LIT, an innovative non-contact thermal characterization method enhanced by tensor analysis. We have developed a straightforward experimental setup based on camera detection, primarily consisting of an IR micro-thermography system and a fiber-coupled laser. The key innovation lies in our unique tensor-based methodology, which leverages the spatially resolved thermal response induced by the heating event. Through detailed analysis and demonstrative measurements, we show that this approach not only substantially improves the quality in determining an arbitrary in-plane thermal conductivity tensor but also achieves this with remarkable efficiency—completing the entire data acquisition process in under 40 seconds, thereby enabling HT data acquisition. Furthermore, our investigation into samples with rough surfaces reveals that SR-LIT exhibits extended robustness in characterizing such samples, accommodating surface roughness up to $r_a = 3.5$ µm. Additional noteworthy features include SR-LIT's ability to handle small samples on the millimeter scale with minimal preparation requirements; a simple surface coating with either metallic transducer or graphite spray is sufficient. Moreover, SR-LIT offers the possibility of measurement for transducerless samples by selecting appropriate pump laser and detector wavelengths.

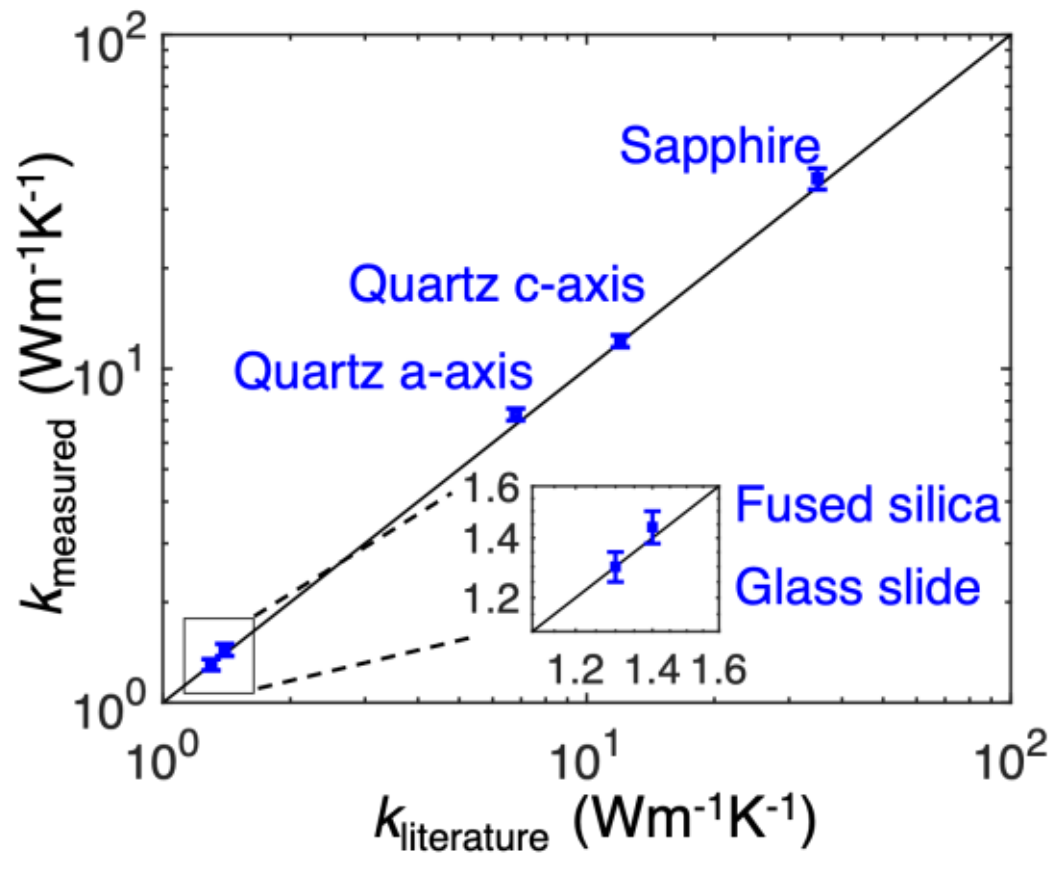

**Figure 11.** Comparison of in-plane thermal conductivities measured using SR-LIT (y-axis) with literature values (x-axis) for fused silica[40], soda-lime glass slide[75], quartz[75], and sapphire[73]. Detailed values can be found in Supplementary Sec. S9.

Looking forward, the versatilities of SR-LIT positions it as an ideal candidate for integration into autonomous robotic experimental platforms. Its potential applications span various domains, including additive manufacturing for real-time monitoring of build properties[83], as well as in autonomous labs[10,36], where its simple preparation requirements and high-throughput thermal characterization offer distinct advantages. Moreover, the tensor-based methodology presents an advanced framework that can seamlessly complement a range of thermometry techniques, such diamond quantum thermometry[84–86], fluorescence thermometry[87–89], and photoacoustic thermometry[90,91], among others. These features collectively render SR-LIT a valuable tool with broad applicability across diverse research areas and real-world applications.

## APPENDIX

### 1. Mathematical model

The 3D heat diffusion equation in cartesian coordinates with no heat-generation term is given by

$$C\frac{\partial T}{\partial t} = k_{xx}\frac{\partial^2 T}{\partial x^2} + k_{yy}\frac{\partial^2 T}{\partial y^2} + k_{zz}\frac{\partial^2 T}{\partial z^2} + 2k_{xy}\frac{\partial^2 T}{\partial x \partial y} + 2k_{xz}\frac{\partial^2 T}{\partial x \partial z} + 2k_{yz}\frac{\partial^2 T}{\partial y \partial z}, \quad (1)$$

where $C$ is the volumetric heat capacity, and $k_{ij}$ with different subscripts are corresponding elements in the thermal conductivity tensor $\boldsymbol{k}$,

$$\boldsymbol{k} = \begin{bmatrix} k_{xx} & k_{xy} & k_{xz} \\ k_{yx} & k_{yy} & k_{yz} \\ k_{zx} & k_{zy} & k_{zz} \end{bmatrix}. \tag{2}$$

Here, $k_{ij} = k_{ji}$ due to the Onsager reciprocal relation[66,67]. The solution for solving the multi-layer 3D heat diffusion model has been well established in the literature[37,92] and can be found in Supplementary Material. Sec. S10. Generally, equation (1) can be transformed into an ordinary differential equation (ODE) by applying the Fourier transform to time $t$ and both in-plane coordinates $x, y$, $T(x, y, z, t) \rightarrow \Theta(u, v, z, \omega)$. The ODE of the multilayered system can then be solved by the thermal quadrupole approach. The resulting surface temperature is

$$\Theta_{\text{top}}(x, y, \omega) = \int_{-\infty}^{\infty} \int_{-\infty}^{\infty} \hat{G}(u, v, \omega) Q_0(u, v, \omega) e^{i2\pi(ux+vy)} du dv, \tag{3}$$

where $Q_0(u, v, \omega)$ is the intensity of absorbed heat flux after the Fourier transform in spatial and time domains, and $\hat{G}(u, v, \omega)$ is the Green's function. Since the heating spot radius and thermal penetration depth are significantly larger than the optical absorption depth, it is reasonable to assume surface absorption[93].

The next step is to determine the thermal response at each pixel, where each pixel is considered a 'probe pixel' with a square shape and constant distribution. Assume the center of the probe pixel is located at $(x_c, y_c)$. Then, the weight function of the probe pixel can be written as

$$I_{\text{probe}}(x_c, y_c) = \int_{x_c - \frac{l_p}{2}}^{x_c + \frac{l_p}{2}} \int_{y_c - \frac{l_p}{2}}^{y_c + \frac{l_p}{2}} \frac{1}{l_p^2} dx dy, \tag{4}$$

where $l_p$ is the side length of the pixel. Finally, the probed thermal response for a pixel at $(x_c, y_c)$ is given by the weighted average of $\tilde{T}_{\text{top}}(x, y)$ by $I_{\text{probe}}(x_c, y_c)$:

$$H(x_c, y_c, \omega) = \frac{1}{l_p^2} \int_{x_c - \frac{l_p}{2}}^{x_c + \frac{l_p}{2}} \int_{y_c - \frac{l_p}{2}}^{y_c + \frac{l_p}{2}} \Theta_{top}(x, y, \omega) dx dy. \tag{5}$$

The thermal response $H(x_c, y_c, \omega)$ has a linear relation with measured data and will be evaluated numerically. A simple and computationally efficient interpretation of the thermal response is that each probe pixel reads the temperature at its center. Therefore, the temperature response is exactly Equation 3.

However, this interpretation might not be accurate when the pixel is large compared to the temperature gradient. Since experimentally we used a reasonably large heating spot size compared to the pixel resolution, it is safe to use this simplification and the accuracy is verified (see Supplementary Material. Sec. S11).

## 2. Sensitivity analysis

Sensitivity analysis is a valuable tool for evaluating the impact of different parameters on the signal and can guide the optimized experimental configuration. Here we use the sensitivity coefficient definition proposed by Gundrum *et al*[94],

$$S^{\gamma} = \frac{\partial \ln(\gamma)}{\partial \ln(\alpha)} \tag{6}$$

where $\gamma$ denotes the choice of signal and $\alpha$ denotes the parameter of interest. Here all input parameters are tested, including metallic transducer properties $k_{xx,m}$, $k_{yy,m}$, $k_{zz,m}$, $C_m$, and transducer thickness $l_m$; substrate properties $k_{xx}$, $k_{yy}$, $k_{zz}$, and $C_s$; and other inputs including interface thermal conductance between the metallic transducer and the substrate $G$, beam spot radius $w_x, w_y$, and pixel side length $l_p$. With a low modulation frequency and small pixel size, we found that all the signals show negligible sensitivity to the film thermal property, $G$, $l_m$, and $l_p$, as such they are omitted in further analysis. The detailed sensitivity analysis is conducted for both in-plane isotropic and in-plane anisotropic materials in Section 4.

## 3. Uncertainty formalism

Our uncertainty formalism follows the work of Yang *et al.*[95] and Seber[96]. This full-error propagation formalism is based on the framework of regression analysis and can be applied to measurements where a known model is fit to one or more observable parameters using a least square algorithm. The loss function of the fitting is defined as:

$$R = \sum_{i=1}^{N} \left[ y_{d_{ofs,i}} - g\left(X_U, X_p, d_{ofs,i}\right) \right]^2, \tag{7}$$

where $y_i$ is the measured signal at the $i$-th offset spot, $g$ is the corresponding value evaluated by the thermal model, $X_U$ is the vector of unknown parameters and $X_p$ is the vector of input parameters. The best-fit unknown parameter will have uncertainty from both experimental noise and uncertainty of input parameters. For the in-plane thermal conductivity fitting, we consider three unknown parameters: $X_U = [k_{xx}, k_{xy}, k_{yy}]^T$, and ten input parameters: $X_{pi} = [k_{xx,m}, k_{yy,m}, k_{zz,m}, C_m, C_s, h_m, l_p, G_s, w_x, w_y]^T$. The resulting variances of unknown parameters are given in the format of variance-covariance matrix:

$$Var[X_U] = \begin{bmatrix} \sigma_{u_1}^2 & cov(u_1, u_2) & \dots \\ cov(u_2, u_1) & \sigma_{u_2}^2 & \cdots \\ \vdots & \vdots & \ddots \end{bmatrix}, \quad (8)$$

where the elements on the principal diagonal are the variance of the unknown parameters, and the value of $\pm 2\sigma$ (95% confidence interval) is used as the uncertainty of corresponding unknown parameters. Detailed derivation can be found in Supplementary Material. Sec. S12. Note that the formula of error-propagation is based on the Taylor series of loss functions, where the derivatives of the signal to the input parameters are computed, and it is similar to the definition of the sensitivity formula. Thus, the sensitivity of the unknown parameter is a good indicator of the corresponding uncertainty, and it is possible to reduce the uncertainty by following the guidance of the sensitivity analysis. Moreover, when considering the error propagation for 2D map-based data, it is interesting to note that the Pearson correlation coefficients $\rho_{k_{xx},k_{yy}}$, $\rho_{k_{xx},k_{xy}}$, and $\rho_{k_{yy},k_{xy}}$ are all close to 1. This indicates a strong linear relation among the unknown parameters[95]. However, the strong linear relation does not affect fitting the three parameters simultaneously, detailed explanation can be found in Supplementary Material. Sec. S13.

# Supplementary: Spatial Resolved Infrared micro-thermography method for thermal conductivity tensor measurement

## S1. Examination of gaussian distributed heating pattern

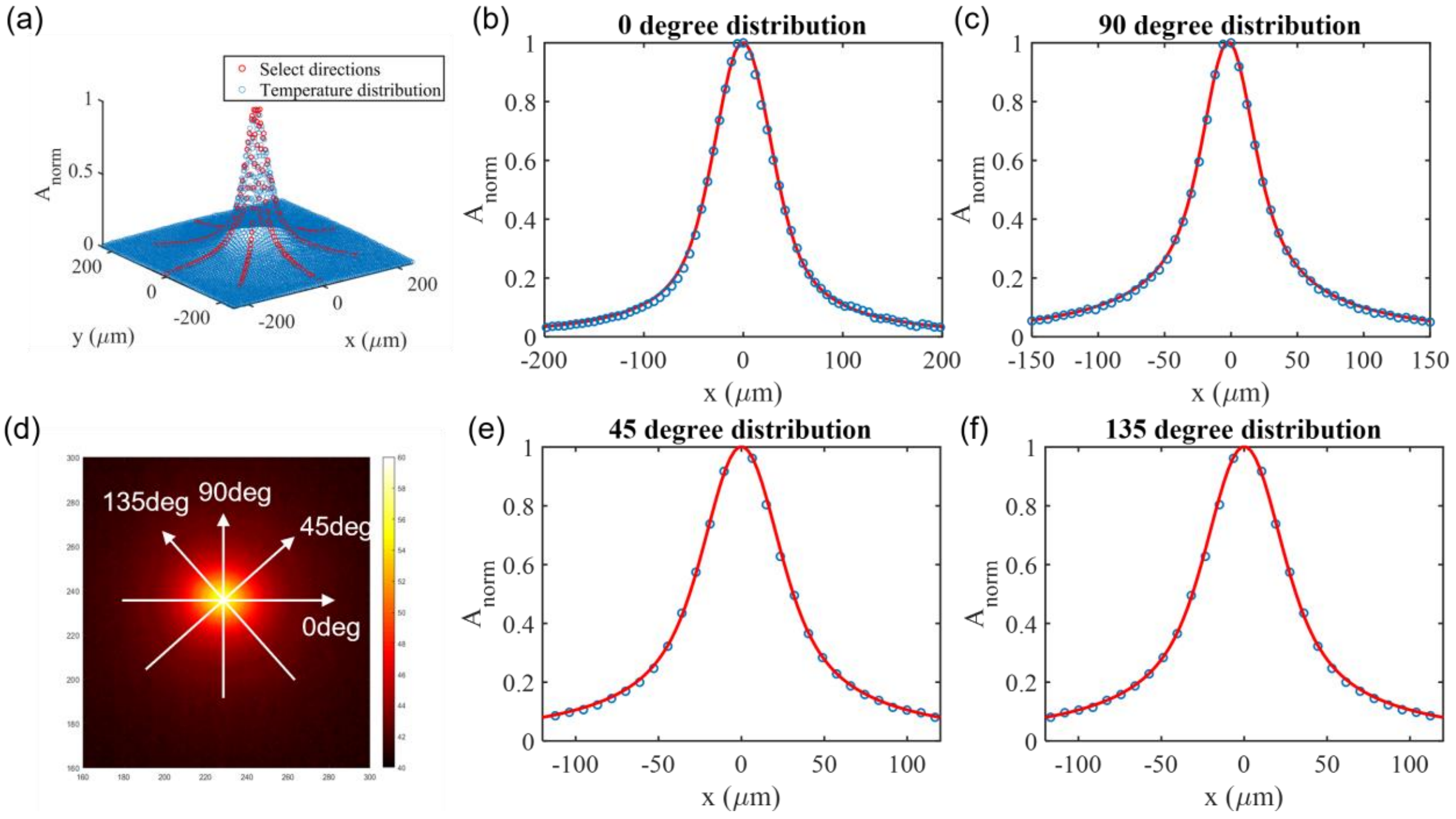

Figure S11 The gaussian beam impinges on the sample with a 30-degree incident angle, creating a slightly elliptical-shaped laser spot. we confirmed that the Gaussian distribution remains consistent, and no peak intensity is observed following the inclination. Although it is possible to directly measure the intensity distribution of the beam by using a beam profiler or adopting knife-edge method, they require additional instruments, and the process is cumbersome. Instead, it is more convenient to check the distribution of the heating-induced temperature response on a sample with isotropic thermal properties, since the distribution only depends on the beam profile. The measurement procedure of the temperature response will be illustrated in main text section 2.3. (a) The normalized temperature response (temperature divided by the maximum temperature at the heating center) on a fused-silica sample, and selected directions for distribution checking, including 0 degree, 45 degree, 90 degree, and 135 degrees. (d) The top view of the temperature response. From the directional plots ((b),(c), (e), and (f)) it can be observed the temperature distribution exhibits perfect Gaussian distribution, as such it is safe to conclude no peak intensity shift occurs after the inclination.

## S2. Emissivity map

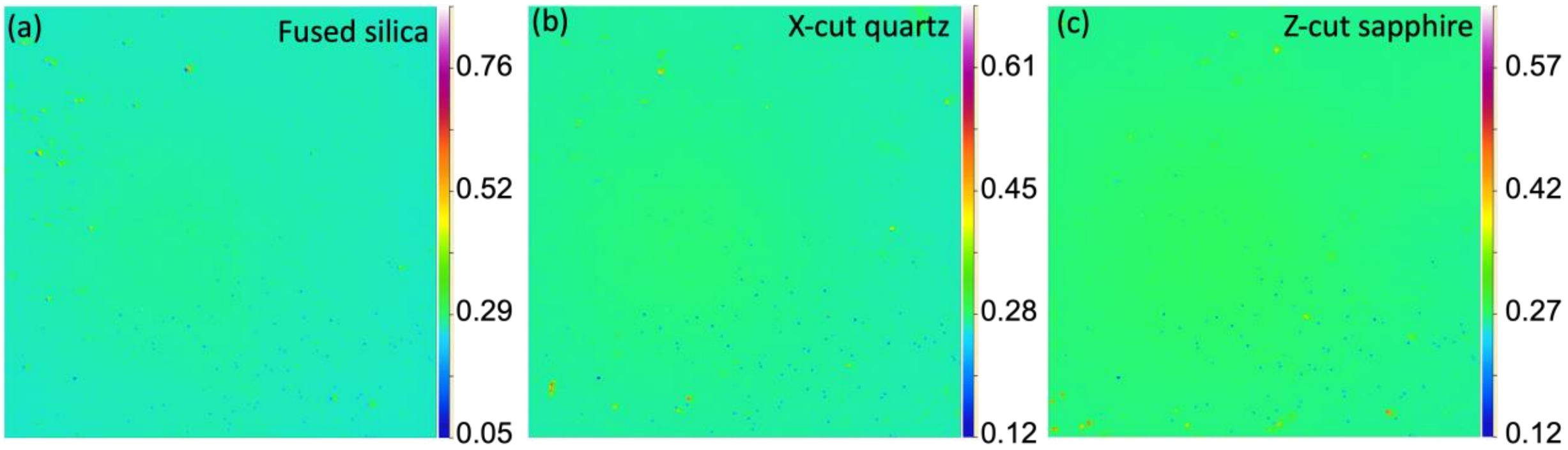

Figure S2 Demonstration of emissivity maps for different samples. (a) Fused silica. (b) x-cut quartz, and (c) z-cut sapphire. Since all samples are coated with a thin Ti transducer layer, the measured emissivity is identical ($\sigma = 0.27$) with deviation smaller than 4%.

The calibration of the emissivity map involves sequentially heating the sample to two elevated temperatures using a thermal stage [1–3]. At each temperature, the corresponding surface IR radiation profile is recorded. Subsequently, by analyzing the surface radiation profiles at the two temperatures, the surface emissivity map $\epsilon_s(x, y)$ can be retrieved.

## S3. Calibration of the tilted angle

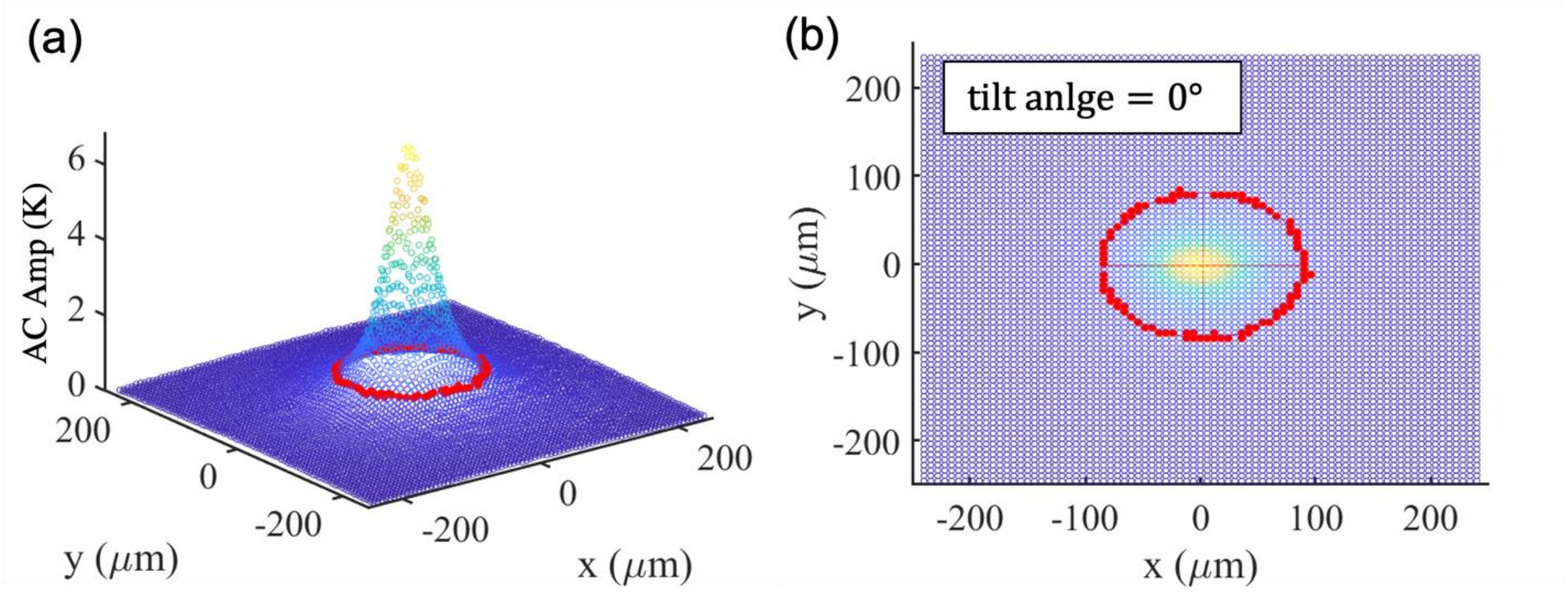

Figure S3 In order to simplify model, the long and the short axis of the heating spot are aligned paralleled to the x- and y-axis of the focus plane array, respectively. The tilted angle can be calibrated by directly fitting an ellipse function fitting. (a) AC amplitude map with $1/e^2$ maximum amplitude ellipse (red dot). (b) The two red lines are the fitted axes of the $1/e^2$ maximum amplitude ellipse, where the tilted angle is found to be 0°. For isotropic material, the distribution of AC amplitude only depends on the heating source. Therefore, by directly fitting the amplitude contour, we can determine the tilted angle of the heating source. However, for transversely anisotropic material, the distribution of AC amplitude also depends on the arrangement of the sample. Thus, for measurement of transversely anisotropic sample with unknown crystal direction, the tilted angle cannot be determined directly. Nevertheless, the tilted angle can be determined by implementing a pre-measurement on an isotropic sample, as such we can guarantee a known tilted angle when implementing subsequent measurement on transversely anisotropic sample.

## S4 FFT Spectrum

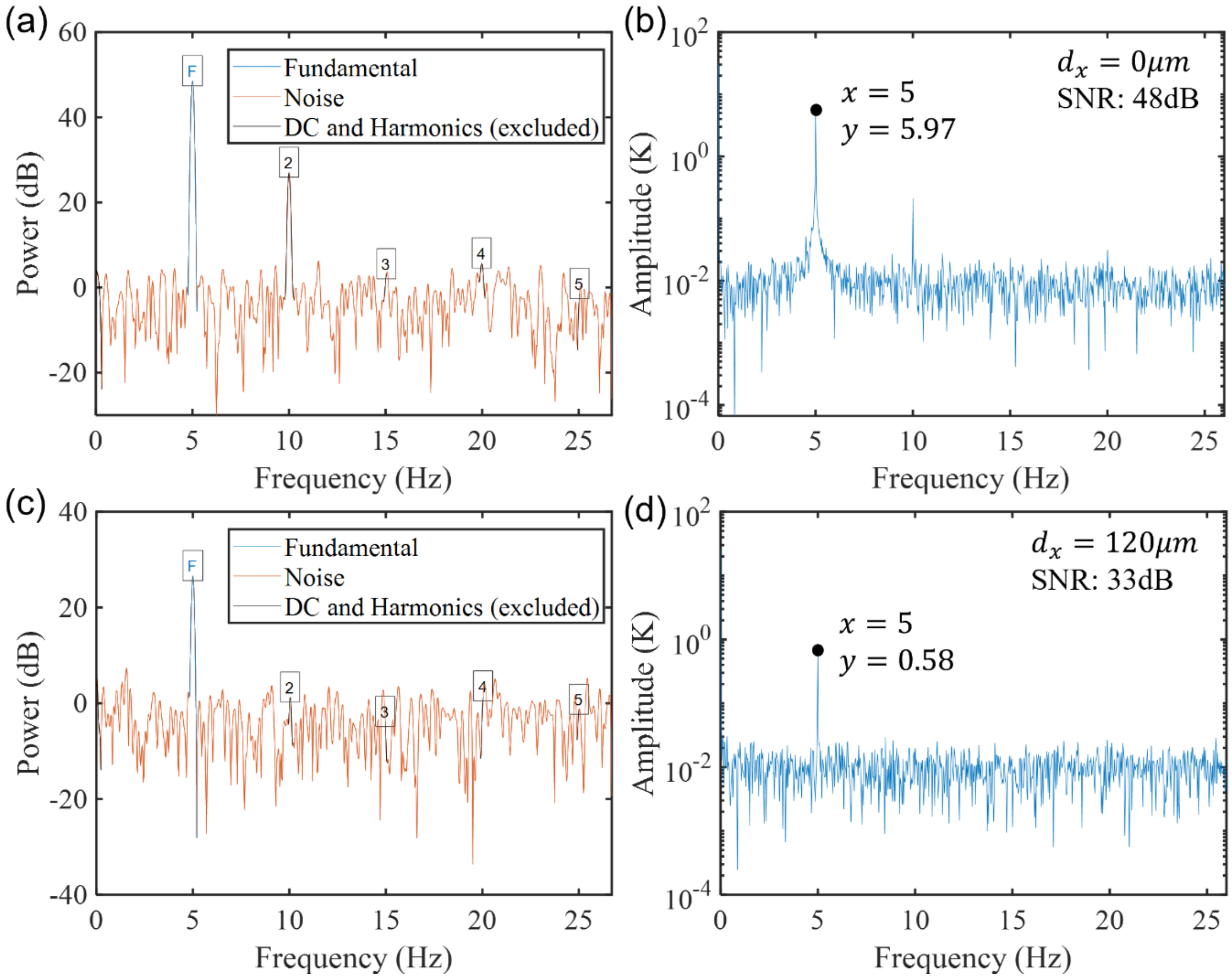

Figure S4 FFT spectrum of measurement of Ti/fused silica sample at modulation frequency 5Hz. The first column represents the power spectrum and the second column represent the Amplitude (Temperature) spectrum. The first row represents the spectrum measured at the heating center and the second row represent the spectrum measured at $d_x = 120$ µm. The pump radii are $w_x = 44.1 \pm 2.4$ µm and $w_y = 25.8 \pm 1.8$ µm. The SNR is defined as the ratio of the peak amplitude to the averaged amplitude of the noise floor in decibels (dB): $SNR_{db} = 20\log_{10}(\frac{A_{peak}}{A_{noise}})$ [15].

In FFT, the $n$ frequency bins are defined as $f_n = {}^{nf_{\mathrm{frame}}}/_{N_{\mathrm{fft}}} = n/t_{\mathrm{frame}}$, where $n$ ranges from 0 to $N_{\mathrm{fft}} - 1$, $f_{\mathrm{frame}}$ is the framerate of the camera, $N_{\mathrm{fft}}$ is the total number of sampling points, and $t_{\mathrm{samp}}$ is the total sampling time. Given that $f_{\mathrm{frame}}$ is fixed, $t_{\mathrm{samp}}$ has a significant impact on the frequency resolution, determining if the signal at the frequency of interest (5 Hz) lands precisely in a frequency bin. We found that for our camera system, the best $t_{\mathrm{samp}}$ is 37 s, which corresponds to $N_{\mathrm{fft}}$=1976.

## S5. Table of nominal values of input parameters

Table S1 Nominal values of input parameters with their uncertainty of 68% confidence interval. The uncertainty of spot radii $w_x, w_y$ are to be determined in the spot radii fitting, details are discussed in main text Appendix. S3.

| Input parameters | Nominal value | Uncertainty (%) |
|---|---|---|
| $w_x$ (Radius of semi-major axis) | 45 μm | NA |
| $w_y$ (Radius of semi-minor axis) | 27 μm | NA |
| $k_f$ (Thermal conductivity of Ti film) | 15 $Wm^{-1}$ $K^{-1}$ | 20 |
| $h_f$ (Thin film thickness) | 100 nm | 5 |
| $l_p$ (Pixel size) | 6 μm | 5 |
| $G_{f/s}$ (Thermal boundary conductance) | 100 MW $m^{-2}K^{-1}$ | 20 |
| $C_f$ (Heat capacity of Ti film) | 2.44 $Jcm^{-3}K^{-1}$ | 2 |
| $C_s$ (Heat capacity of substrate) | 1.62 $Jcm^{-3}K^{-1}$ | 2 |

# S6 Average Filter

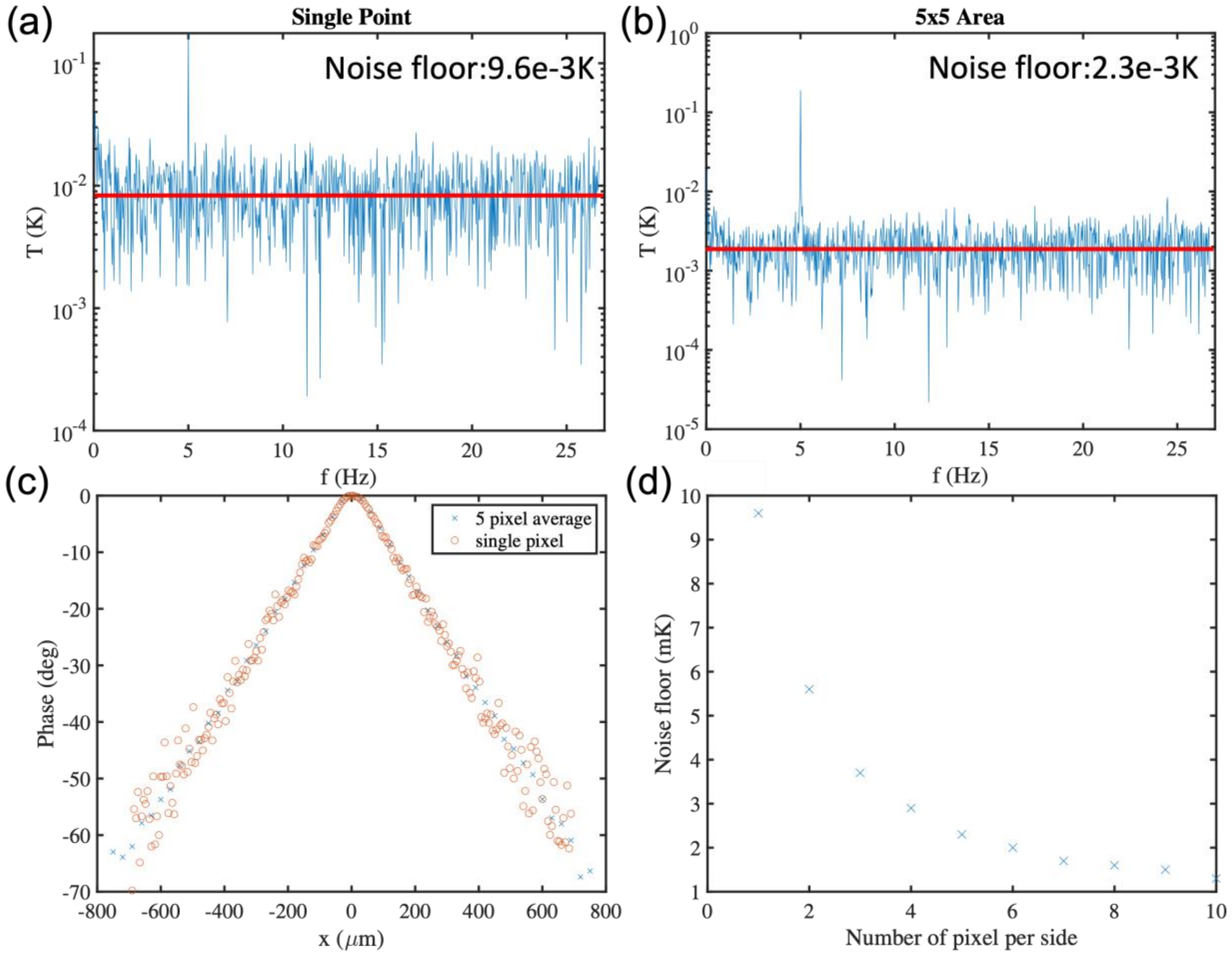

Figure S5 Exemplar average filter for denoising. (a) FFT spectrum before average filter, where the noise floor is $9.6e^{-3}$ K. (b) FFT spectrum after applying a $5 \times 5$ filter, the noise floor is reduced to $2.3e^{-3}$ K. (c) Phase as a function of offset distance $x$. The noise in the phase of a single pixel is significantly larger that of the average filter. (d) Noise floor as a function of size of average filter, from $1 \times 1$ (no filter) to $10 \times 10$. The noise floor decreases with the increase in the size of the average filter. Noted that the spatial resolution is compromised with the increasing size of the average filter, a balance needs to be found between the filter size and the spatial resolution. A window with size $5 \times 5$ to $7 \times 7$ reaches the balance.

## S7 Sensitivity map for in-plane anisotropic thermal conductivity tensor

In this section, a detailed sensitivity analysis is given. In-plane anisotropic materials can be further divided into two categories based on the arrangement of their in-plane lattice vectors: material with orthogonal in-plane lattice vectors (i.e., $\vec{c} \perp \vec{a}$), such as x- and y-cut quartz; and material with non-orthogonal in-plane lattice vectors, such as (010) $\beta - \mathrm{Ga_2O_3}$, where the axial angle between [001] and [100] is $103.7°$. For materials having orthogonal in-plane lattice vectors, there are two experiment conditions to consider: 1. Samples with known crystal direction and 2. Samples with unknown crystal direction. In the first case, where the sample has known crystal direction, it is possible to arrange the sample in such a way that the two in-plane lattice vectors, $\vec{c}$ and $\vec{a}$, align with the *x*- and *y*-axes of the optical system, respectively (see the configuration sketch in Figure S6 (a)). Subsequently, the in-plane thermal conductivity tensor based on the optical coordinate system, often referred to principle in-plane thermal conductivity tensor, can be written as $\boldsymbol{k_{in,1}} = \begin{bmatrix} k_{x_1x_1} & 0 \\ 0 & k_{y_1y_1} \end{bmatrix} = \begin{bmatrix} k_c & 0 \\ 0 & k_a \end{bmatrix}$, where $k_c$ and $k_a$ are the lattice thermal conductivity along $\vec{c}$ and $\vec{a}$ direction, respectively.

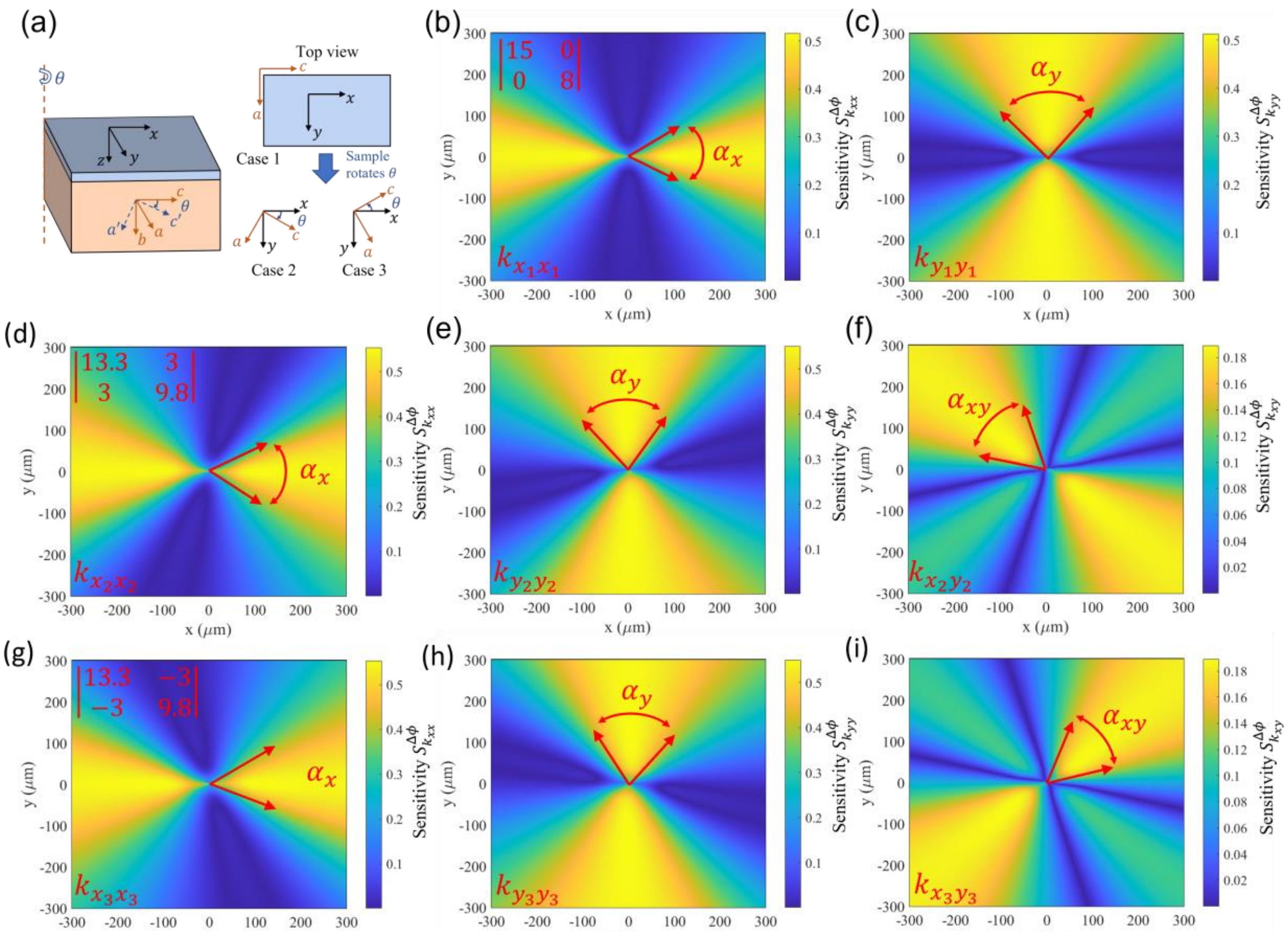

Figure S6 The $\Delta\phi$ sensitivity map for $k_{xx}, k_{yy}$, and $k_{xy}$. (a) Schematic for 3 different sample configurations. In case 1 (first row), $\vec{c}$ and $\vec{a}$ are aligned parallel to the x- and y-axes of the optical system, respectively; In case 2 and case 3 (second and third row), the sample is intentionally rotated $30°$ counterclockwise and clockwise, respectively. Note that since the optical system has positive z direction goes into the paper, the counterclockwise rotation about z axis appears to be clockwise from top view. (b), (d) and (g): $\Delta\phi$ Sensitivity map for $k_{xx}$, where $\alpha_x$ is the region that $k_{xx}$ exhibits high sensitivity; (c), (e) and (h): $\Delta\phi$ Sensitivity map for $k_{yy}$, with high sensitivity region $\alpha_y$; And last, (f) and (i): $\Delta\phi$ Sensitivity map for $k_{xy}$ with high sensitivity region $\alpha_{xy}$. All input parameters are same as listed in Table.1, and the specific in-plane thermal conductivity tensor used for each row is printed on $\Delta\phi$ Sensitivity map.

Figure S6 (b) and (c) present the sensitivity map of $\Delta\phi$ to $k_{xx}$ and $k_{yy}$ based on a hypothetical in-plane thermal conductivity tensor $\boldsymbol{k_{in,1}} = \begin{bmatrix} 15 & 0 \\ 0 & 8 \end{bmatrix}$. Similar to the sensitivity map for isotropic materials, $\Delta\phi$ exhibits high sensitivity to $k_{xx}$ and $k_{yy}$ in a region within an offset angle $\alpha_x, \alpha_y$. However, in contrast to isotropic materials, $\alpha_x$ for anisotropic thermal conductivity tensor is narrower, while $\alpha_y$ is broader. This indicates the ratio of $k_{xx}$ and $k_{yy}$ affects the influence region of $k_{xx}$ and $k_{yy}$.

In the second case, $\vec{c}$ and $\vec{a}$ cannot be aligned parallel to the x- and y-axes since the crystal direction is unknown. Therefore, the principle in-plane thermal conductivity tensor $\boldsymbol{k_{in,2}} = \begin{bmatrix} k_{x_2x_2} & k_{x_2y_2} \\ k_{y_2x_2} & k_{y_2y_2} \end{bmatrix}$, has non-zero off-diagonal terms, and $k_{x_2x_2} \neq k_c$, $k_{y_2y_2} \neq k_a$, and $k_{x_1y_1} = k_{y_1x_1} \neq 0$. Assuming we are examining the same hypothetical material as case 1, and the sample is counterclockwise rotated $30°$ about z axis, then the to-be-determined $\boldsymbol{k_{in,2}} = \boldsymbol{Q^T k_{in,1} Q} = \begin{bmatrix} 13.3 & 3 \\ 3 & 9.8 \end{bmatrix}$, where $\boldsymbol{Q}$ is the transformation matrix $\boldsymbol{Q} = \begin{bmatrix} cos\theta & sin\theta \\ -sin\theta & cos\theta \end{bmatrix}$4. The positive thermal conductivity of $k_{xy}$ suggest that the temperature gradient in the $x$ direction would lead to a heat flux in the positive $y$ direction, and vice versa. In line with the observation in case 1, $\Delta\phi$ exhibits high sensitivity to $k_{xx}$ and $k_{yy}$ within $\alpha_x$ and $\alpha_y$ , respectively, but with a slightly twisted opening orientation (Figure S6 (d) and (e)). Furthermore, the off-diagonal term $k_{xy}$ shows high sensitivity in a direction pointing to $135°$ and within the offset angle $\alpha_{xy}$ (Figure S6 (f)). Remarkably, the absolute sensitivity value for $k_{xy}$ is comparable to that of $k_{xx}$ and $k_{yy}$. Given the comparable sensitivity levels and the distinct sensitivity patterns for each in-plane thermal conductivity tensor element, simultaneously fitting all the three elements becomes feasible. In addition, after fitting $\boldsymbol{k_{in,2}}$, the rotation angle $\theta$ can be deduced. The rotation angle satisfies the relation: $k_{x_1y_1} = 0 = k_{x_2y_2}\cos(\theta) - \frac{k_{x_2x_2}\sin(2\theta)}{2} + \frac{k_{y_2y_2}\sin(2\theta)}{2}$, as such $\theta = \frac{1}{2}\arctan\left(\frac{2k_{x_2y_2}}{k_{x_2x_2} - k_{y_2y_2}}\right)$. Finally, for materials with non-orthogonal in-plane lattice vectors, the sensitivity pattern is generally the same as this second case since they both have the non-zero $k_{xy}$. However, for these materials, the rotation angle cannot be directly retrieved using the same relation since the crystalline in-plane thermal conductivity tensor itself already contains a non-zero $k_{xy}$ component.

Same as mentioned in main text, it is important to recognize that the sensitivity map pattern for each element is not fixed. Consider a case 3 where the hypothetical tensor $\boldsymbol{k_{in,1}}$ is rotated clockwise by 30°, then the to-be-determined principle in-plane thermal conductivity tensor becomes $\boldsymbol{k_{in,3}} = \begin{bmatrix} 13.3 & -3 \\ -3 & 9.8 \end{bmatrix}$. In this case the high sensitivity region for $k_{xy}$ is now pointing towards 45°, which is distinctly different from case 2 (Figure S6 (i)).

# S8. Sample with rough surface

## S8.1 Far-field phase with different shape of heating spot

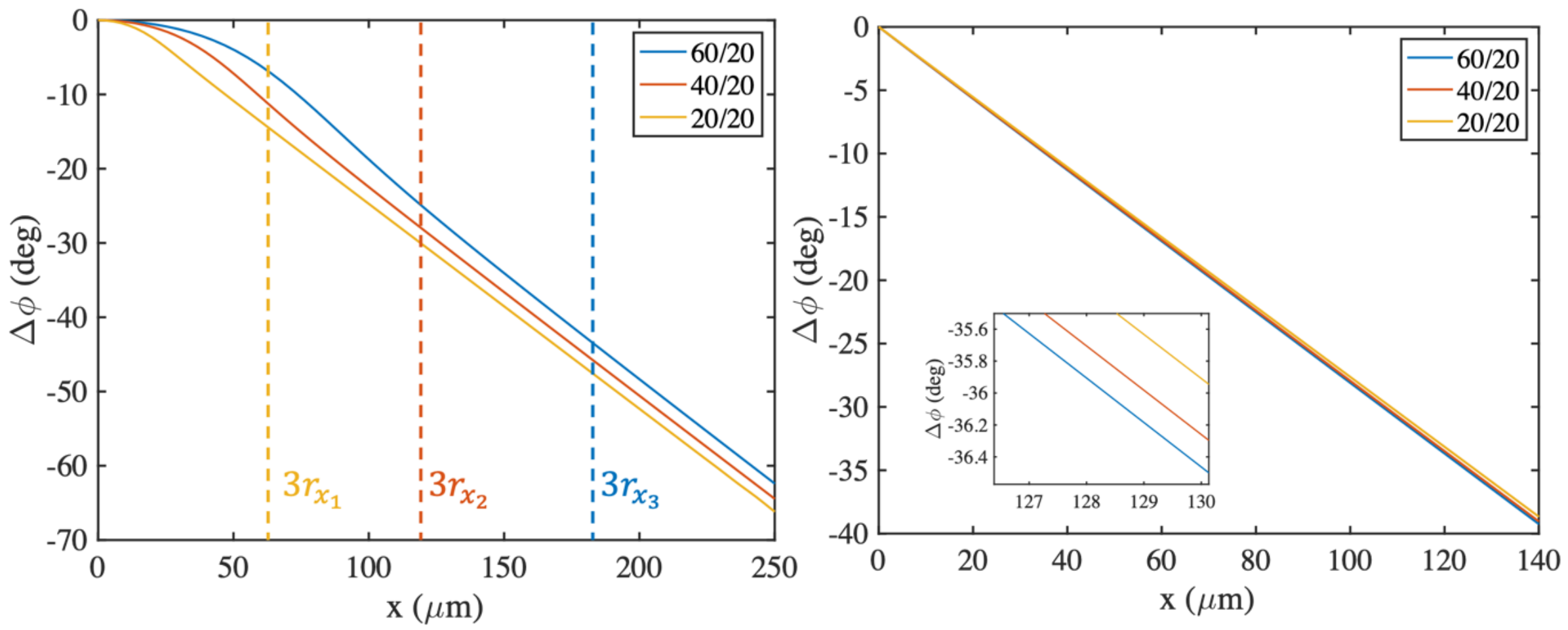

Figure S7 Compare far-field $\Delta\phi$ with different shape of heating spot. Blue curve: $r_x = 60$ μm, $r_y = 20$ μm; Orange curve: $r_x = 40$ μm, $r_y = 20$ μm; Yellow curve: $r_x = 20$ μm, $r_y = 20$ μm. (a) plots $\Delta\phi$ starts from heating center. (b) plots $\Delta\phi$ starts from the far-field (cut-off position from (a)). The difference among the far-field ($x > 3r_x$) phase for different heating spot shape is smaller than 1%, therefore it is safe to ignore the shape of the heating spot when analyzing the far-field data.

## S8.2 Best-fit map for rough samples

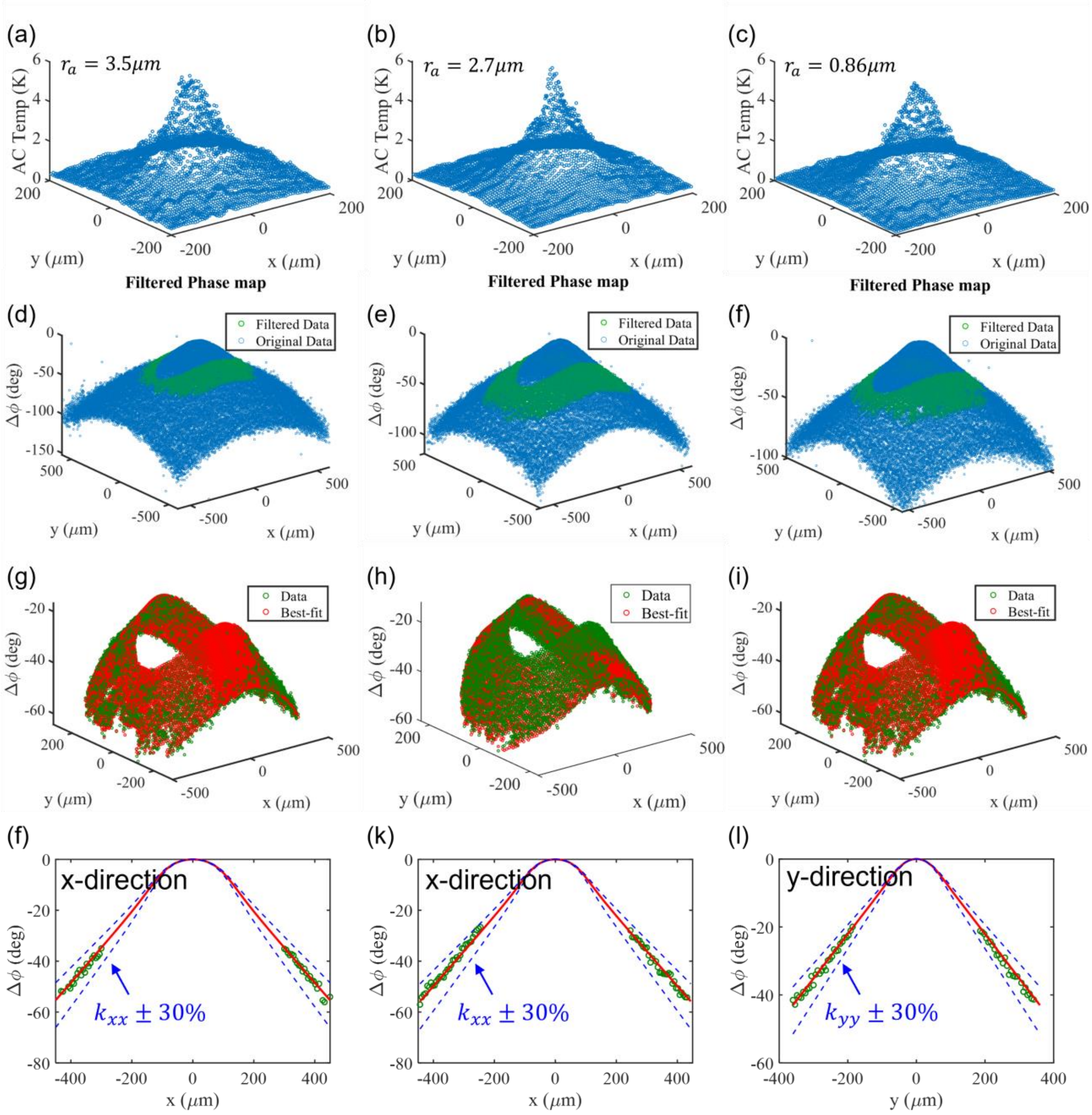

Figure S8 Rough glass slide measurement data maps. (a)-(c) AC amplitude map. (d)-(f) Measured $\Delta\phi$ map with filtered region. (g)-(i) modeled $\Delta\phi$ map based on best-fit thermal conductivity tensor with measured $\Delta\phi$ map. (f)-(i) 1D $\Delta\phi$ visualization. The roughness $r_a$ are 3.5 μm, 2.7 μm, and 0.86 μm, from left to right.

### S8.3 Rough sample measurement results

Table.S2 The in-plane thermal conductivities of samples with poor surface finish. The $\pm$uncertainties represent 95% confidence interval.

| Sample | $\boldsymbol{k_{xx}}(\mathbf{W/m \cdot K})$ | $\boldsymbol{k_{yy}}(\mathbf{W/m \cdot K})$ |
|---|---|---|
| Soda-lime glass | | |
| Pristine ($\boldsymbol{R_a} = \mathbf{3\ nm}$) | $\mathbf{1.31 \pm 0.07}$ | $\mathbf{1.34 \pm 0.06}$ |
| $\boldsymbol{R_a} = \mathbf{228\ nm}$ | $\mathbf{1.34 \pm 0.09}$ | $\mathbf{1.33 \pm 0.06}$ |
| $\boldsymbol{R_a} = \mathbf{0.86\ \mu m}$ | $\mathbf{1.38 \pm 0.07}$ | $\mathbf{1.36 \pm 0.06}$ |
| $\boldsymbol{R_a} = \mathbf{2.7\ \mu m}$ | $\mathbf{1.41 \pm 0.05}$ | $\mathbf{1.34 \pm 0.06}$ |
| $\boldsymbol{R_a} = \mathbf{3.5\ \mu m}$ | $\mathbf{1.4 \pm 0.07}$ | $\mathbf{1.36 \pm 0.06}$ |
| x-cut quartz | | |
| Pristine | $\mathbf{12.1 \pm 0.5}$ | $\mathbf{7.3 \pm 0.4}$ |
| $\boldsymbol{R_a} = \mathbf{1.5\ \mu m}$ | $\mathbf{12.3 \pm 0.6}$ | $\mathbf{7.3 \pm 0.3}$ |

## S9. Summary of all measurement materials

Table.S3 Summary of in-plane thermal conductivities of transversely isotropic materials measured by SR-IRT and the literature values. The $\pm$uncertainties represent 95% confidence interval.

| Sample | Literature | | | | Measurement | | |
|---|---|---|---|---|---|---|---|
| | $k_{xx}$ (W/m · K) | $k_{yy}$ (W/m · K) | $k_{xy}$ (W/m · K) | $C_v$(J/cm$^3$ · K) | $k_{xx}$ (W/m · K) | $k_{yy}$ (W/m · K) | $k_{xy}$ (W/m · K) |
| Fused silica | 1.3~1.5[5,6] | 1.3~1.5[5,6] | - | 1.62[5] | 1.48 $\pm$ 0.06 | 1.5 $\pm$ 0.06 | - |
| Soda-lime glass | 1.3[5,7,8] | 1.3[5,7,8] | - | 2.17[9] | 1.31 $\pm$ 0.1 | 1.34 $\pm$ 0.1 | - |
| (0001) Sapphire | 35[10–12] | 35[10–12] | - | 3.06[13] | 37.7 $\pm$ 2.7 | 37.2 $\pm$ 2.6 | - |
| x-cut quartz (Aligned) | 12[8] | 6.8[8] | 0 | 1.98[8] | 12.1 $\pm$ 0.5 | 7.2 $\pm$ 0.3 | 0.03 $\pm$ 0.1 |
| x-cut quartz (**20°** rotation) | 11.4 | 7.41 | 1.67 | 1.98 | 11.2 $\pm$ 0.5 | 7.6 $\pm$ 0.4 | 1.6 $\pm$ 0.14 |

## S10. 3D Anisotropic heat transport model

### S10.1 Heat diffusion in a multilayered system with anisotropic thermal conductivities

Here a general case of a multilayer system is considered in the thermal model, each layer with homogeneous but anisotropic thermal conductivities. The governing equation of the heat diffusion is:

$$C\frac{\partial T}{\partial t}=k_x\frac{\partial^2 T}{\partial x^2}+k_y\frac{\partial^2 T}{\partial y^2}+k_z\frac{\partial^2 T}{\partial z^2}+2k_{xy}\frac{\partial^2 T}{\partial x\partial y}+2k_{xz}\frac{\partial^2 T}{\partial x\partial z}+2k_{yz}\frac{\partial^2 T}{\partial y\partial z}\quad . \tag{S10-1}$$

This parabolic partial differential equation can be simplified into ODE by doing Fourier transforms with respect to in-plane coordinates and time, $T(\mathrm{x,y,z,t})\leftrightarrow\Theta(\mathrm{u,v,z},\omega)$

$$F(u)=\int_{-\infty}^{\infty}f(x)e^{-i2\pi ux}dx$$

$$\mathrm{F}\left\{\frac{df(x)}{dx}\right\}=i2\pi uF(u)$$

$$\mathrm{F}\left\{\frac{d^2f(x)}{dx^2}\right\}=-\left(2\pi u\right)^2F(u)$$

$$\left(iC\omega\right)\Theta=-4\pi^2\left(k_xu^2+2k_{xy}uv+k_yv^2\right)\Theta+2i2\pi\left(k_{xz}u+k_{yz}v\right)\frac{\partial\Theta}{\partial z}+k_z\frac{\partial^2\Theta}{\partial z^2}, \tag{S10-2}$$

or more compactly,

$$\frac{\partial^2\Theta}{\partial z^2}+\lambda_2\frac{\partial\Theta}{\partial z}-\lambda_1\Theta=0\,. \tag{S10-3}$$

Where

$$\lambda_1\equiv\frac{iC\omega}{k_z}+\frac{4\pi^2\left(k_xu^2+2k_{xy}uv+k_yv^2\right)}{k_z}, \tag{S10-4}$$

$$\lambda_2\equiv 2i2\pi\frac{\left(k_{xz}u+k_{yz}v\right)}{k_z}\,. \tag{S10-5}$$

The general solution of Eq. (S3-3) is

$$\Theta=e^{u^+z}B^++e^{u^-z}B^-, \tag{S10-6}$$

where $u^+,u^-$ are the roots of the equation $x^2+\lambda_2x-\lambda_1=0$:

$$u^{\pm} = \frac{-\lambda_2 \pm \sqrt{(\lambda_2)^2 + 4\lambda_1}}{2}, \tag{S10-7}$$

and $B^{+}, B^{-}$ are the complex numbers to be determined.

The heat flux can be obtained from the temperature Eq. (S3-6) and Fourier's law of heat conduction $Q = -k_z \left( d\Theta / dz \right)$ as:

$$Q = -k_z u^{+} e^{u^{+} z} B^{+} - k_z u^{-} e^{u^{-} z} B^{-}. \tag{S10-8}$$

It would be convenient to write Eq. (S3-6) and (S3-8) as matrices as

$$\begin{bmatrix} \Theta \\ Q \end{bmatrix}_{n,z} = [N]_n \begin{bmatrix} B^{+} \\ B^{-} \end{bmatrix}_n. \tag{S10-9}$$

$$[N]_n = \begin{bmatrix} 1 & 1 \\ -k_z u^{+} & -k_z u^{-} \end{bmatrix} \begin{bmatrix} e^{u^{+} z} & 0 \\ 0 & e^{u^{-} z} \end{bmatrix}_n, \tag{S10-10}$$

where *n* stands for the *n*-th layer of the multilayer system, and *z* is the distance from the surface of the *n*-th layer.

The constants $B^{+}, B^{-}$ for the *n*-th layer can also be obtained from the surface temperature and heat flux of that layer by setting *z*=0 in Eq. (S3-10) and performing its matrix inversion:

$$\begin{bmatrix} B^{+} \\ B^{-} \end{bmatrix}_n = [M]_n \begin{bmatrix} \Theta \\ Q \end{bmatrix}_{n,z=0}, \tag{S10-11}$$

$$[M]_n = \frac{1}{k_z \left( u^{+} - u^{-} \right)} \begin{bmatrix} -k_z u^{-} & -1 \\ k_z u^{+} & 1 \end{bmatrix}. \tag{S10-12}$$

For heat flow across the interface, an interface conductance G is defined. Therefore, the heat flux across an interface can be written as:

$$Q_{n,z=L} = Q_{n+1,z=0} = G\left( \Theta_{n,z=L} - \Theta_{n+1,z=0} \right). \tag{S10-13}$$

From Eq. (S3-13) we also have:

$$\Theta_{n+1,z=0} = \Theta_{n,z=L} - \frac{1}{G} Q_{n,z=L}. \tag{S10-14}$$

It is convenient to write Eq. (S3-13) and (S3-14) collectively as a matrix,

$$\begin{bmatrix}\Theta \\ Q\end{bmatrix}_{n+1,z=0} = [R]\begin{bmatrix}\Theta \\ Q\end{bmatrix}_{n,z=L}, \tag{S10-15}$$

$$[R] = \begin{bmatrix}1 & -1/G \\ 0 & 1\end{bmatrix}. \tag{S10-16}$$

The surface temperature and heat flux can thus be related to those at the bottom of the substrate as

$$\begin{bmatrix}\Theta \\ Q\end{bmatrix}_{\mathrm{n},z=L_n} = [N]_n[M]_n \cdots [R]_1[N]_1[M]_1\begin{bmatrix}\Theta \\ Q\end{bmatrix}_{1,z=0} = \begin{bmatrix}A & B \\ C & D\end{bmatrix}\begin{bmatrix}\Theta \\ Q\end{bmatrix}_{1,z=0}. \tag{S10-17}$$

For the modeling of suspended thin films, a semi-infinite substrate composed of air can be added at the bottom. The boundary condition of zero heat flux at the bottom of the substrate thus yielding $0 = C\Theta_{1,z=0} + DQ_{1,z=0}$. The Green's function $\hat{G}$, which is essentially the detected temperature response due to the applied heat flux of unit strength, can thus be solved as

$$\hat{G}(u,v) = \frac{\Theta_{1,z=0}}{Q_{1,z=0}} = -\frac{D}{C}. \tag{S10-18}$$

With the Green's function $\hat{G}$ determined, the detected temperature response is simply the product of $\hat{G}$ and the heat source function in the frequency domain. See details in the next sub-section.

For the case with surface heat loss, the matrix is

$$\begin{bmatrix}\Theta \\ Q\end{bmatrix}_{\mathrm{n},z=L_n} = [N]_n[M]_n \mathrm{L}\ [R]_1[N]_1[M]_1\begin{bmatrix}\Theta \\ Q - Q_{loss}\end{bmatrix}_{1,z=0} = \begin{bmatrix}A & B \\ C & D\end{bmatrix}\begin{bmatrix}\Theta \\ Q - U\Theta\end{bmatrix}_{1,z=0}. \tag{S10-19}$$

Where $Q_{loss} = U\Theta$, with $U$ [W/m$^2$K] as the heat loss coefficient. Applying the same boundary condition of zero heat flux at the bottom of the substrate yields $0 = C\Theta_{1,z=0} + D(Q_{1,z=0} - U\Theta_{1,z=0})$. The Green's function $\hat{G}$ can thus be solved as

$$\hat{G}(u,v) = \frac{\Theta_{1,z=0}}{Q_{1,z=0}} = -\frac{D}{C - UD}. \tag{S10-20}$$

A focused continuous wave laser beam having a Gaussian distribution of intensity over space $p_0(x, y)$ and modulated by a sinusoidal function is used to heat up the sample:

$$p_0(x,y,t)=\frac{2A_0}{\pi\sigma_{x_0}\sigma_{y_0}}\exp\left(-\frac{2x^2}{\sigma_{x_0}^2}\right)\exp\left(-\frac{2y^2}{\sigma_{y_0}^2}\right)e^{i\omega_0 t}. \tag{S10-21}$$

Where $A_0$ is the average power of the pump beam; $\sigma_{x_0}$ and $\sigma_{y_0}$ are the $1/e^2$ radii of the pump spot in the $x$ and $y$ directions, respectively; $w_0$ is the modulation frequency. Only the periodic heat input is considered here, as the DC component will be removed by the lock-in amplifier. Fourier transform of $p_0(x,y,t)$ over the space and time is

$$P_0(u,v,\omega)=A_0\exp\left(-\frac{\pi^2u^2\sigma_{x_0}^2}{2}\right)\exp\left(-\frac{\pi^2v^2\sigma_{y_0}^2}{2}\right)2\pi\delta(\omega-\omega_0). \tag{S10-22}$$

Here we have utilized the following relationship

$$\mathcal{F}\{e^{-ax^2}\}=\int_{-\infty}^{\infty}e^{-ax^2}e^{-i2\pi ux}dx=\sqrt{\left(\frac{\pi}{a}\right)}e^{-\pi^2u^2/a}. \tag{S10-23}$$

The detected temperature response is the product of the surface heat flux $P_0(u,v,\omega)$ and the Green's function $\hat{G}(u,v,\omega)$ in the frequency domain. Inverse Fourier transform yields the real space surface temperature distribution as

$$\theta(x,y,\omega)=\int_{-\infty}^{\infty}\int_{-\infty}^{\infty}P_0(u,v,\omega)\hat{G}(u,v,\omega)e^{i2\pi(ux+vy)}dudv. \tag{S10-24}$$

## S10.2 Modeling of signals acquired in the experiments

The next step is to determine the thermal response at each pixel, where each pixel is considered a 'probe pixel' with a cube shape and constant distribution. Assume the center of the probe pixel locates at $(x_c,y_c)$ Then, the weight function of the probe pixel can be written as

$$I_{probe}(x_c,y_c)=\int_{x_c-\frac{l_p}{2}}^{x_c+\frac{l_p}{2}}\int_{y_c-\frac{l_p}{2}}^{y_c+\frac{l_p}{2}}\frac{1}{l_p^2}dxdy, \tag{S10-25}$$

Where $l_p$ is the side length of the pixel. Finally, the probed thermal response for pixel at $(x_c,y_c)$ is given by the weighted average of $\tilde{T}_{top}(x,y)$ by $I_{probe}(x_c,y_c)$:

$$H(x_c, y_c, \omega) = \frac{1}{l_p^2} \int_{x_c - \frac{l_p}{2}}^{x_c + \frac{l_p}{2}} \int_{y_c - \frac{l_p}{2}}^{y_c + \frac{l_p}{2}} \theta(x, y, \omega) dx dy. \tag{S10-26}$$

$H(x_c, y_c, \omega)$ in Eq. (S22) is a complex number, with the real part as the in-phase output and the imaginary part as the out-of-phase output of the lock-in amplifier:

$$X = Re\{H\}, Y = Im\{H\}. \tag{S10-27}$$

The thermal response $H(x_c, y_c, \omega)$ has a linear relation with measured data and will be evaluated numerically. The definition of integration grid can be found in our previous work. A simple and computationally efficient interpretation of the thermal response is that each probe pixel reads the temperature at its center. Therefore, the temperature response is exactly Equation 3. However, this interpretation might not be accurate when the pixel is large compared to the temperature gradient. Since experimentally we used a reasonably large heating spot size compared to the pixel resolution, it is safe to use this simplification and the accuracy is verified in the following section.

## S11. Comparison between exact pixel reading and point simplification

Computing the exact solution based on Eq. S10-22 can be time-consuming. However, if the pixel is small enough compared to the temperature gradient, it is reasonable to adopt a simplification that the pixel reads the temperature at its center. Therefore, the temperature response simplifies to Eq.3 in the main text and the computational time is two orders of magnitude faster. Here we investigate when the simplification is valid. Assume a material with a small thermal conductivity $k = 1.4\ \mathrm{Wm^{-1}K^{-1}}$ and $C_v = 1.6\ \mathrm{W\ cm^{-3}\ K^{-1}}$, so that the thermal diffusion length is as small as possible. With experiment condition of pixel size of 6 μm, and modulation frequency of 5 Hz. The normalized amplitude and relative phase under different heating spot size is shown in figure below:

For cases where the heating spot size is smaller or comparable to the probe pixel (as shown in Figure. S8 (a),(b),(e),(f)), the relative phase calculated using a simplified solution shows a slight offset compared to the exact solution, and there are more notable differences in the normalized amplitude. However, when the spot size is larger than the pixel size (as illustrated in Figure. S8 (c),(d),(g),(h)), the difference between the simplified and exact solutions becomes negligible. Given that the smallest heating spot size in actual experiment is around 26 μm, which is four times larger than the probe pixel, it is reasonable to utilize the simplified model to achieve better computational efficiency without compromising accuracy.

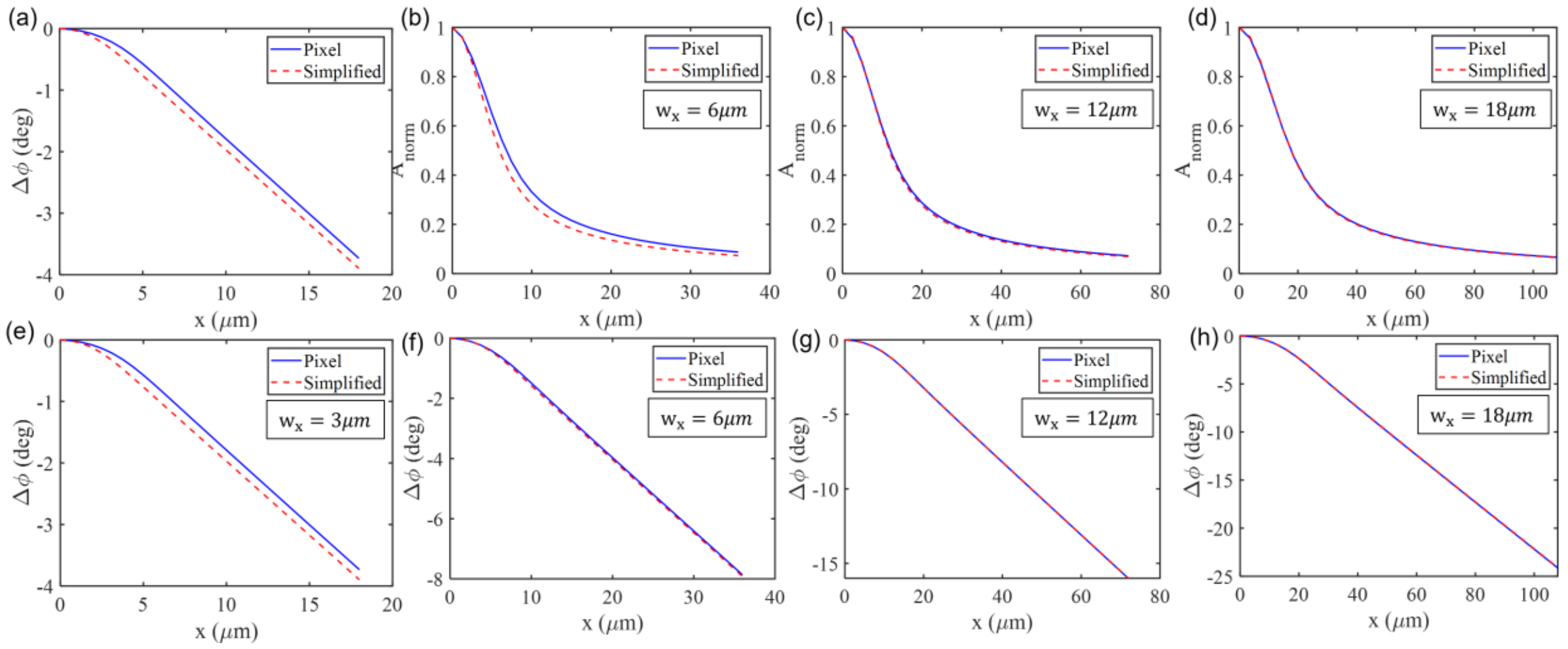

Figure S9 Normalized amplitude and relative phase as a function of offset distance with different spot sizes. Each column corresponds to a different heating spot size, ranging from left to right: 3 μm, 6 μm, 12 μm, and 18 μm. The first row represents the normalized amplitude, and the second row represents the relative phase. The probe pixel size is 6 μm.

## S12. Uncertainty formalism

In processing the data, we extract multiple parameters ($k_{xx}$,$k_{yy}$ and $k_{xy}$) from experimental data simultaneously, using the least-squared regression method. This process could be mathematically expressed as seeking to minimize the squared difference between the experimental data and the model predictions[14]:

$$R(\vec{X}_U) = \sum_{i=1}^{N}\left[y_i - g\left(\vec{X}_U, \vec{X}_P, d_i\right)\right]^2, \quad \text{(S12-1)}$$

where $y_i$ is the $i\,th$ measured signal at offset spot $d_i$, $g$ is the corresponding value evaluated by the thermal model, $\vec{X}_U$ is the vector of unknown parameters and $\vec{X}_P$ is the vector of controlled parameters. For in-plane thermal conductivity tensor fitting, we consider total three unknown parameters:

$\vec{X}_U = [k_{xx}, k_{xy}, k_{yy}]^T$, and ten controlled parameters $\vec{X}_P = [k_{xx,f}, k_{yy,f}, f, C_f, C_s, h_f, l_p, G_s, w_x, w_y]^T$. Note that since in computation it is safe to assume that the metallic transducer has isotropic thermal conductivity, i.e., $k_{xx,f} = k_{yy,f} = k_{zz,f}$, since the thickness of the transducer is significantly smaller than the thermal characteristics length. At the best fit, the gradient of $R$ should be zero for every element in $\vec{X}_U$:

$$\frac{\partial R(\vec{X}_U)}{\partial x_u}\bigg|_{\hat{\vec{X}}_U} = \sum_{i=1}^{N}\left[\left(y_i - g(\vec{X}_U, \vec{X}_P, d_i)\right)\frac{\partial g(\vec{X}_U, \vec{X}_{\boldsymbol{P}}, d_i)}{\partial x_u}\bigg|_{\hat{\vec{X}}_{\boldsymbol{U}}}\right] = 0, \quad (u = 1,2,\dots,n_u) \tag{S12-2}$$

where $\hat{\vec{X}}_{\boldsymbol{U}}$ is the least squares estimate of the unknown parameter vector $\vec{X}_U$. Denote the mean values of $\hat{\vec{X}}_{\boldsymbol{U}}$, $\vec{X}_P$ as $\vec{X}_U^*$, $\vec{X}_P^*$, near the small neighbor of $\vec{X}_U^*$, $\vec{X}_P^*$, $g(\vec{X}_U, \vec{X}_P, d_i)$ can be rewritten by applying Taylor expansion,

$$g(\vec{X}_U, \vec{X}_P, d_i) \approx g(\vec{X}_U^*, \vec{X}_P^*, d_i) + \sum_{u=1}^{n_u}\frac{\delta g(\vec{X}_U, \vec{X}_P, d_i)}{\partial x_u}\bigg|_{\vec{X}_U^*, \vec{X}_P^*}(x_u - x_u^*) + \sum_{p=n_u+1}^{M}\frac{\partial g(\vec{X}_U, \vec{X}_P, d_i)}{\partial x_p}\bigg|_{\vec{X}_U^*, \vec{X}_P^*}(x_p - x_p^*). \quad i = 1,2,\dots,N \tag{S12-3}$$

Substitute Eq. (S12-3) into Eq. (S12-2),

$$\sum_{i=1}^{N}\left[\left(y_i - g(\vec{X}_U^*, \vec{X}_P^*, d_i) - \sum_{u=1}^{n_u}\frac{\partial g(\vec{X}_U, \vec{X}_P, d_i)}{\partial x_u}\bigg|_{\vec{X}_U^*, \vec{X}_P^*}(\hat{x}_u - x_u^*) - \sum_{p=1}^{n_p}\frac{\partial g(\vec{X}_U, \vec{X}_P, d_i)}{\partial x_p}\bigg|_{\vec{X}_U^*, \vec{X}_P^*}(x_p - x_p^*)\right) \times \frac{\partial g(\vec{X}_U, \vec{X}_P, d_i)}{\partial x_u}\bigg|_{\vec{X}_U^*, \vec{X}_P^*}\right] = 0. \tag{S12-4}$$

Eq. (S12-4) can be re-written into matrix form:

$$\boldsymbol{J}_{\boldsymbol{U}}^{*\prime}\left[\vec{y} - \vec{g}(\vec{X}_U^*, \vec{X}_P^*, d_i) - \boldsymbol{J}_{\boldsymbol{U}}^*\left(\hat{\vec{X}}_{\boldsymbol{U}} - \vec{X}_U^*\right) - \boldsymbol{J}_{\boldsymbol{P}}^*(\vec{X}_P - \vec{X}_P^*)\right] = \vec{0}_{p\times 1}, \tag{S12-5}$$

where $\boldsymbol{J}_{\boldsymbol{U}}^*$, $\boldsymbol{J}_{\boldsymbol{P}}^*$ are the Jacobian matrix:

$$\boldsymbol{J_U^*} = \begin{bmatrix} \frac{\partial g(\vec{X}_U,\vec{X}_P,d_1)}{\partial x_1} & \cdots & \frac{\partial g(\vec{X}_U,\vec{X}_P,d_1)}{\partial x_{n_u}} \\ \vdots & \ddots & \vdots \\ \frac{\partial g(\vec{X}_U,\vec{X}_P,d_N)}{\partial x_1} & \cdots & \frac{\partial f_n(\theta)}{\partial x_{n_u}} \end{bmatrix}_{N\times n_u}, \qquad \text{(S12-6)}$$

$$\boldsymbol{J_P^*} = \begin{bmatrix} \frac{\partial g(\vec{X}_U,\vec{X}_P,d_1)}{\partial x_{n_u+1}} & \cdots & \frac{\partial g(\vec{X}_U,\vec{X}_P,d_1)}{\partial x_M} \\ \vdots & \ddots & \vdots \\ \frac{\partial g(\vec{X}_U,\vec{X}_P,d_N)}{\partial x_{n_u+1}} & \cdots & \frac{\partial f_n(\theta)}{\partial x_M} \end{bmatrix}_{N\times n_p}. \qquad \text{(S12-7)}$$

Here we utilize notation $x_i\ (i = 1,2,\ \ldots, n_u) = x_u\ (u = 1,2,\ \ldots, n_u)$, and $x_i\ (i = n_u + 1, n_u + 2, \ldots, M) = x_p(p = 1,2,\ \ldots, n_p)$, where $n_u$ is total number of unknown parameters, $n_p$ is the total number of controlled parameters, $M = n_u + n_p$ is the total number of parameters. If $\boldsymbol{J_U^*}'\boldsymbol{J_U^*}$ is an invertible matrix, then Eq. (S12-5) can be rearranged to form:

$$\hat{\vec{X}}_{\boldsymbol{U}} = (\boldsymbol{J_U^*}'\boldsymbol{J_U^*})^{-1}\boldsymbol{J_U^*}'\left[\vec{y} - \vec{g}\left(\vec{X}_U^*, \vec{X}_P^*, d_i\right) - \boldsymbol{J_P^*}\left(\vec{X}_P - \vec{X}_P^*\right)\right] + \vec{X}_U^*. \qquad \text{(S12-8)}$$

Assuming the experimental noise at each spot has a normal distribution, and the controlled parameters are also normally distributed around their mean values, the variance-covariance matrix of $\hat{\vec{X}}_{\boldsymbol{U}}$ then is:

$$Var(\hat{\vec{X}}_{\boldsymbol{U}}) = (\boldsymbol{J_U^*}'\boldsymbol{J_U^*})^{-1}\boldsymbol{J_U^*}'\left[Var(\vec{y}) - \vec{g}\left(\vec{X}_U^*, \vec{X}_P^*, d_i\right) - J_P^* Var(\vec{X}_P) J_P^{*\prime}\right]\boldsymbol{J_U^*}(\boldsymbol{J_U^*}'\boldsymbol{J_U^*})^{-1}. \qquad \text{(S12-9)}$$

Where $Var(\vec{y})$ and $Var(\vec{X}_P)$ are the variance matrix of the experiment measurement and the variance matrix of controlled parameters, respectively:

$$Var(\vec{y}) = \begin{bmatrix} \sigma_{y_1}^2 & \cdots & 0 \\ \vdots & \ddots & \vdots \\ 0 & \cdots & \sigma_{y_N}^2 \end{bmatrix}_{N\times N}, \qquad \text{(S12-10)}$$

$$Var\left(\vec{X}_P\right) = \begin{bmatrix} \sigma_{x_{c_1}}^2 & \cdots & 0 \\ \vdots & \ddots & \vdots \\ 0 & \cdots & \sigma_{x_{c_{n_p}}}^2 \end{bmatrix}_{n_p\times n_p}. \qquad \text{(S12-11)}$$

$\sigma_{y_i}^2, i = 1,2,\dots,N$, are the variances of experiment measured data at spot $y_i$; $\sigma_{x_{c_i}}^2, i = 1,2,\dots,n_p$, are the variances of controlled parameter $x_{c_i}$. Therefore, the uncertainty emerges from both experimental noise and controlled parameters are all considered. Finally, the uncertainties of the unknown parameters can be determined form the diagonal terms of Eq. (S12-9):

$$Var(\hat{\vec{X}}_{\boldsymbol{U}}) = \begin{pmatrix} \sigma_{x_{u_1}}^2 & cov[x_{u_1}, x_{u_2}] & cov[x_{u_1}, x_{u_3}] & \\ cov[x_{u_2}, x_{u_1}] & \sigma_{x_{u_2}}^2 & cov[x_{u_2}, x_{u_3}] & \cdots \\ cov[x_{u_3}, x_{u_1}] & cov[x_{u_3}, x_{u_2}] & \sigma_{x_{u_3}}^2 & \\ & \vdots & & \ddots \end{pmatrix}. \quad \text{(S12-12)}$$

Eq. (S12-10) is the same result as illustrated in main text Eq. 8. Moreover, it is interesting to note that the Pearson correlation coefficients $\rho_{k_{xx},k_{yy}}$, $\rho_{k_{xx},k_{xy}}$, and $\rho_{k_{yy},k_{xy}}$ are all close to 1, which indicates a strong linear relation among the unknown parameters. However, the strong linear relation does not affect fitting the three parameters simultaneously, detailed explanation can be found in supplementary material. Sec. S13.

## S13. Pearson correlation coefficient (PCC) and loss function

### S13.1 Pearson correlation coefficient

Table. S4 Pearson correlation coefficient and corresponding uncertainty (96% confidence interval). The uncertainty of $k_{xy}$ based on x-axis and y-axis data is not shown since the extremely large value.

| Data set | Pearson correlation coefficient | | | Uncertainty (%) | | |
|---|---|---|---|---|---|---|
| | $\rho_{k_{xx},k_{yy}}$ | $\rho_{k_{xx},k_{xy}}$ | $\rho_{k_{yy},k_{xy}}$ | $k_{xx}$ | $k_{yy}$ | $k_{xy}$ |
| Map fitting | 0.96 | 0.98 | 0.99 | 6 | 5 | 13 |
| Directional data (x-axis data only) | 0.89 | 0.97 | 0.89 | 98 | 255 | - |
| Directional data (y-axis data only) | 0.74 | 0.74 | 0.98 | 469 | 108 | - |
| Directional data (4 directions) | 0.96 | 0.89 | 0.88 | 14.22 | 12.9 | 60 |

The Pearson correlation coefficient (PCC) is defined as: $r_{i,j} = \frac{cov[x_{u_i}, x_{u_j}]}{\sigma_{x_{u_i}}\sigma_{x_{u_j}}}$, where $cov[x_{u_i}, x_{u_j}]$ is the covariance of $x_{u_i}$and $x_{u_j}$, $\sigma_{x_{u_i}}$ is the variance of $x_{u_i}$. PCC quantifies strength of linear dependence between two parameters. It is a value between -1 and 1, inclusive, the closer the PCC is to the boundary, the stronger of the linear dependence between two parameters is. The PCC can be computed conveniently based on the variance-covariance matrix $Var(\hat{\vec{X}}_{\boldsymbol{U}})$ given in Eq. (S12-10). Here $Var(\hat{\vec{X}}_{\boldsymbol{U}})$is computed based on the hypothetical sample with thermal conductivity tensor given in Section 4.2.2: $k_{in} =$

$\begin{bmatrix} 20 & 2 & 0 \\ 2 & 8 & 0 \\ 0 & 0 & 8 \end{bmatrix}$, with $Var(\vec{y})$ substituted by a constant variance $(0.1°)$ to eliminate biased correlation coefficient. As mentioned in the main text, the in-plane thermal conductivity tensor is retrieved by both map fitting and directional fitting for comparison. For the directional fitting, we consider cases: 1. Only x-axis data. 2. Only y-axis data. And 3. 4 directions data evenly distributed from -90 to 90 degree (i.e., ,0,45,90,135).

It can be found that for map fitting, the Pearson correlation coefficient (PCC) $\rho_{k_{xx},k_{yy}}$, $\rho_{k_{xx},k_{xy}}$, and $\rho_{k_{yy},k_{xy}}$ are all close to 1, which indicates a strong linear relation among the unknown parameters. Subsequently, one might raise questions about the fitting quality and uncertainty of the result, since the common understanding is that the high correlation coefficient leads to increased uncertainty in the result.

However, this statement does not necessarily imply that the fitting quality will be compromised in a nonlinear fitting scenario. Instead, the nonlinear least squares fitting approach has the capability to capture more complex relationships beyond linear associations, as demonstrated by the convex loss function with a single global minimum observed in the plot of the loss function as a function of $k_{xx}$ and $k_{yy}$ with fixed $k_{xy}$ (Figure. S10 (a)). In this specific $\Delta\phi$ map fitting problem, the strong linear correlation among $k_{xx}$, $k_{yy}$, and $k_{xy}$ indicates fixed in-plane anisotropic ratios. Therefore, even in the presence of strong linear correlation, the nonlinear model is still capable of fitting the three tensor elements simultaneously with good accuracy and low uncertainty. On the counterpart, when examining data only offset along x-axis with $\rho_{1,2}$ is well below 1, the corresponding cost function appears to be non-convex (Figure. S10 (b) and (c)), and the uncertainty to $k_{yy}$ is extremely high indicating $k_{yy}$ cannot be fitted simultaneously with $k_{xx}$.

## S13.2. Loss function plots

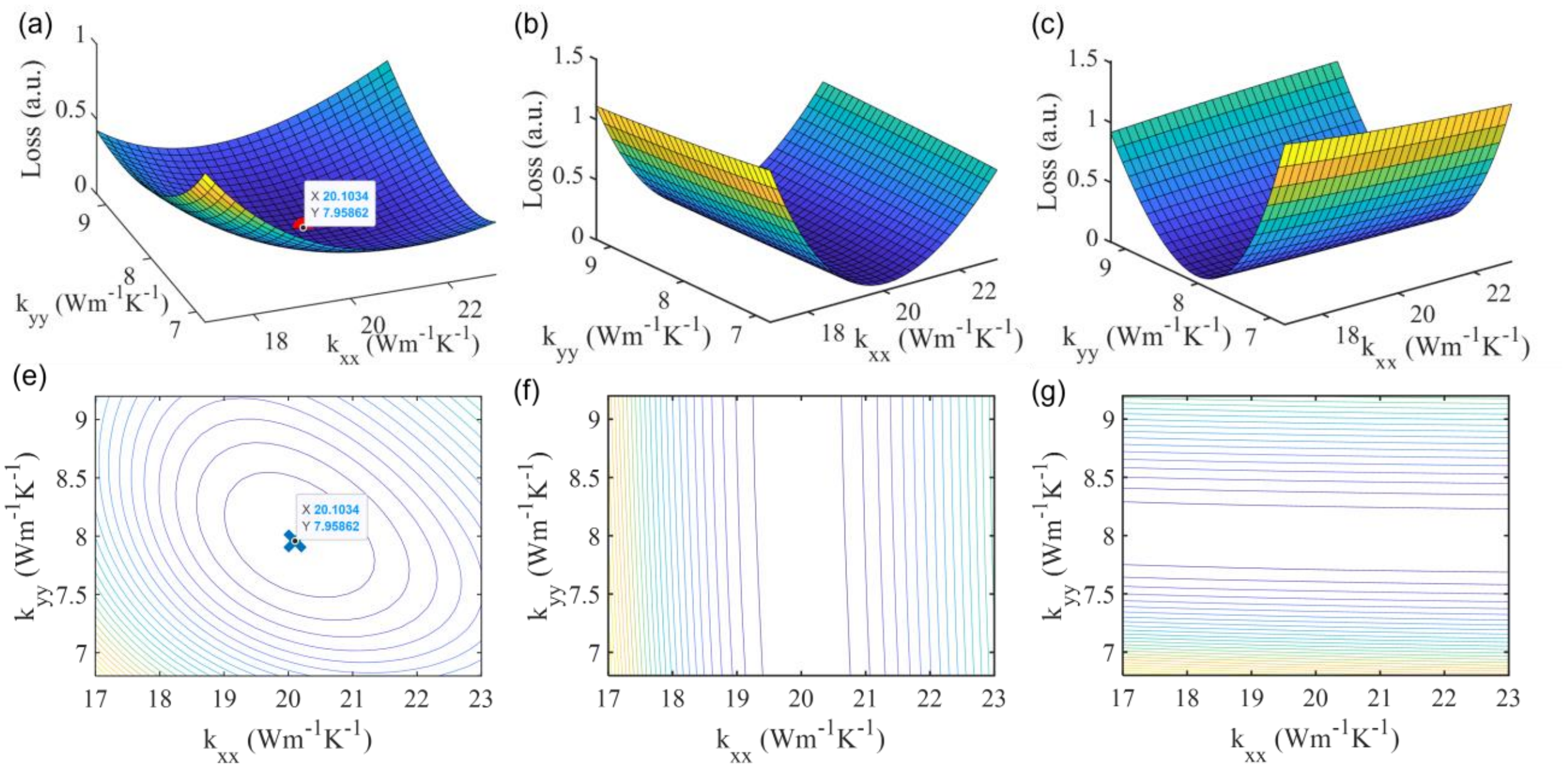

Figure S10 (a)-(c) Loss function map of different choices of data and (e)-(f) corresponding contour plots. The loss function is computed based on fixed $k_{xy} = 2\ \mathrm{Wm^{-1}K^{-1}}$, corresponds to the thermal conductivity tensor introduced in main text section 4.2 $\boldsymbol{k} = \begin{bmatrix} 20 & 2 & 0 \\ 2 & 8 & 0 \\ 0 & 0 & 8 \end{bmatrix}$. The loss function in first column is computed based on map data, where we can clearly observe that the loss function has a convex shape. The second (b) and third (c) loss functions are computed based on solely x-axis and y-axis data, respectively. Those loss functions exhibit non-convex shape with low sensitivity to the tensor element on the orthogonal direction. Therefore, the fitting results are not ideal.